\documentclass[aps,floats,twocolumn,prb,showpacs]{revtex4-2}
\usepackage[utf8]{inputenc}   
\usepackage[T1]{fontenc}      
\usepackage[english]{babel}   
\usepackage{lmodern}          

\usepackage{amsmath}
\usepackage{graphicx}
\usepackage{xcolor}
\usepackage{epstopdf}
\usepackage{natbib}
\usepackage{xfrac}
\usepackage[normalem]{ulem}
\usepackage{multirow}
\usepackage{physics}
\usepackage{amssymb}
\usepackage{lipsum}
\usepackage{verbatim}
\usepackage{hyperref}
\hypersetup{colorlinks=true,urlcolor=blue,citecolor=blue,linkcolor=blue,breaklinks=true}

\newcommand{\old}[1] {}

\begin{document}
\title{Comparing scattering rates from Boltzmann  and dynamical mean-field theory}


\author{M.~Wais$^1$}
\thanks{These two authors contributed equally.} 
\author{J. Kaufmann$^1$}
\thanks{These two authors contributed equally.}
\author{M. Battiato$^2$}
\author{K. Held$^1$}
\affiliation{$^1$ Institute of Solid State Physics, TU Wien,  Vienna, Austria}
\affiliation{$^2$ Nanyang Technological University, 21 Nanyang Link, Singapore, Singapore}

\begin{abstract}
  We compute scattering rates for electrons in the two-dimensional Hubbard model for a one-orbital metal and a two-orbital band insulator
  by means of the  Boltzmann scattering equation (BSE) and dynamical mean-field theory (DMFT). 
  As an intermediate method between both, we also consider the BSE without momentum conservation.
  In the weak interaction regime and for the band insulator, the last two agree to very good accuracy. The BSE with momentum conservation, on the other hand, shows slightly larger scattering rates, and a momentum differentiation of these on the Fermi surface. For the Mott insulator at strong interaction, the  DMFT electron scattering rates are much larger and defy a BSE description.   
 Noteworthy, the scattering rates for the band insulator are exceedingly small because---in contrast to the Mott insulator---there is virtually no impact ionization.
  \end{abstract}

\date{\today}

\maketitle

\section{Introduction}
The electronic structure, i.e., the electronic states and their broadening or scattering rate, is arguably the most fundamental property of a solid.
Scattering processes not only affect equilibrium properties but are also essential  if a  material is driven away from equilibrium. Experimentally, the one-particle scattering rate for the (occupied) electronic states can be measured by angular-resolved photoemission spectroscopy (ARPES)\cite{Grioni2001,Damascelli2003}.
If vertex corrections can be neglected, there is a one-to-one correspondence between this one-particle scattering rate 
and the two-particle scattering rates for response functions such as the
optical conductivity. Here, the width of the Drude peak corresponds to the two-particle scattering rate that, 
without vertex corrections, is directly related to the one-particle scattering rates we calculate here\cite{Drude1900,Drude1900a} \footnote{Both in dynamical mean-field theory and Boltzmann there are no vertex corrections to the optical conductivity}. 
We study them by using two methods that are widely employed in solid state theory, albeit by different communities. 
Through our comparison, we hope to contribute to a better mutual understanding of the strengths and weaknesses of these methods, 
as well as of the very different electron-electron scattering in a metal, band insulator, and Mott insulator.

Dynamical mean-field theory (DMFT) \cite{Metzner1989,Georges1992a,Jarrell1992,Georges1996} 
is one of the most widely used approaches for strongly correlated materials. 
It is non-perturbative and maps a correlated lattice model onto the solution 
of an Anderson impurity model in a self-consistent way \cite{Georges1992a}. 
DMFT becomes exact in the limit of high dimensions or a high connectivity of the lattice \cite{Metzner1989}, 
which implies that the self-energy and hence the scattering rate is momentum independent.

The Boltzmann scattering equation (BSE) \cite{Boltzmann1872,Snoke2007,Ziman1960,Chambers1990} 
has been originally developed for gases \cite{Boltzmann1872} but is nowadays used to address a multitude of different problems, 
all the way from nuclear physics to  cosmology. Often the transport part of this equation is combined with a 
crude approximation for the scattering, the relaxation time approximation,  to study transport properties. 
However, the full Boltzmann scattering term can also be included, allowing e.g.\ for a highly detailed 
reconstruction of the thermalization process. Among the possible applications of the full Boltzmann scattering 
term is the possibility of calculating scattering rates, making a direct comparison of this approach with DMFT possible.

To the best of our knowledge such comparison has not been done in a systematic way, 
and we attempt to fill this blank spot through  this work. 
Specifically, we study the equilibrium scattering rates for the single-orbital Hubbard model 
in two dimensions as well as those for a two-orbital band insulator. 
The BSE is expected to fail at strong interaction $U$, since it describes the dynamics of the  distribution function 
by a (momentum-resolved) rate equation with 
the transition rates usually calculated in lowest order perturbation theory in $U$ (Fermi's golden rule).
DMFT, on the other hand, neglects (as an impurity model is solved) 
the momentum dependence of scattering (an approximation known to become correct in the limit of high dimensions). 
A final point is that, while DMFT even allows for the construction of effective (local) scattering matrix elements 
in form of the two-particle vertex, the Boltzmann scattering term needs them as input 
and only performs the joint density of states (DOS) integration and,
eventually, the time propagation.

In this paper, we show that indeed at strong interaction $U$, i.e., in the Mott insulating phase \footnote{For an overview of  the Mott-Hubbard transition and the physics of the Mott insulator, see \cite{Gebhard1997}.}, 
a BSE description of the scattering rate is not possible. This is surprising given a good description of the spectral 
redistribution caused by impact ionization \cite{wais2018}. 
The DMFT scattering rate is much higher than what can be expected or understood in a rigid band picture; 
it is intimately connected with the formation of the Hubbard bands and shoulders therein.
Conversely, at weak $U$ we obtain a discrepancy as well. These discrepancies, noticeable larger scattering rates 
and a momentum differentiation on the Fermi surface, can be traced back to the momentum conservation or lack thereof: 
DMFT and  BSE without  momentum conservation are in good agreement.

This paper is structured as follows: 
In Sec.\ \ref{sec:model-methods} we introduce the Hubbard-type models considered, and
describe how scattering rates are calculated in DMFT and with the Boltzmann scattering equation.
In Sec.\ \ref{chap:weak1band} we present results for the weak-coupling
single-orbital Hubbard model. 
Next, we compare scattering rates for the two-orbital band insulator in Sec.\ \ref{weakCoupling2Band} 
and the  Mott insulating single-orbital Hubbard model in Sec.\ \ref{sec:mott}.
In Sec.\ \ref{sec:conclusion} we summarize the results.
Furthermore we provide additional derivations and results in the Appendix.

\section{Model and methods}\label{sec:model-methods}

\subsection{Hubbard-type models}
In this paper we study the single-orbital Hubbard model on a two-dimensional square lattice, 
as well as a related two-orbital model which is a band insulator.
It is useful to employ second quantization,
where operators $c^{(\dag)}_{\mathbf{k}m\sigma}$ 
annihilate (create) electrons at momentum $\mathbf{k}$ and spin $\sigma$ in orbital $m$. 
Their Fourier-transformed operators $c^{(\dag)}_{i m\sigma}$ do the same for a lattice site $i$ 
instead of momentum $\mathbf{k}$;
the products $n^{\vphantom{\dag}}_{\mathbf{k}m\sigma}=c_{\mathbf{k}m\sigma}^\dag c^{\vphantom{\dag}}_{\mathbf{k}m\sigma}$
and $n^{\vphantom{\dag}}_{im\sigma}=c_{im\sigma}^\dag c^{\vphantom{\dag}}_{im\sigma}$
are the particle number operators for momentum and site occupations, respectively. 
Both Hubbard-type models can be described by the following Hamiltonian
\begin{equation}
  \label{eq:hubbard-hamiltonian}
  H = \sum_{\bold k m\sigma} 
    \epsilon^{\vphantom{\dag}}_{m}(\mathbf{k}) 
      n^{\vphantom{\dag}}_{\mathbf{k}m\sigma}
    + \frac{U}{2}  \sum_{i} \! \sum_{(l \sigma) \neq(m \sigma')}\! n_{il\sigma} n_{im\sigma'}.
\end{equation}
The first term constitutes a tight-binding description of the system.
It describes the kinetic energy (``hopping'') of non-interacting electrons
with crystal momentum $\mathbf{k}$ and 
a dispersion relation $\epsilon^{\vphantom{\dag}}_m(\mathbf{k})$
that is assumed to be diagonal in the orbital index $m$. This term is diagonal in momentum space.

The second term models the Coulomb repulsion $U$ between electrons.
It is strictly local at each lattice site $i$ and, for the sake of simplicity,
we take the interaction to be  the same within one orbital and between different orbitals.
Consistently, there is no Hund's rule coupling, i.\ e.\ $J=0$. A self-interaction is excluded in the sum. 
We  consider both, the prevalent single-orbital Hubbard model, where orbital indices $m$ and $l$ are restricted to this single orbital, 
and a two-orbital band insulator with interaction $U$. 
In the latter case, the bandgap is encoded in the dependence of $\epsilon^{\vphantom{\dag}}_m(\mathbf{k})$ on $m\in\{1,2\}$ as detailed below.

Due to the exponential scaling of the Fock space needed
to represent an $N$-particle wave function,
it is completely impossible to compute the dynamics of every
single electron in the system. Instead one is bound to make approximations
such as the DMFT and BSE, for extracting relevant information from
statistically averaged quantities such as distributions or correlation functions.

\subsection{Dynamical mean field theory}
Many-body quantum field theory, which also is the pillar upon which DMFT is built, 
has the Green's function as its basic one-particle quantity. 
The retarded Green's function is defined as follows (with operators in the Heisenberg representation) \cite{Abrikosov1975a}:
\begin{align}
  &G_R(\mathbf{k}, m, t) = -i\Theta(t)\Big\langle c_{\mathbf{k}m \sigma}^{\vphantom{\dag}}(t) c_{\mathbf{k}m \sigma}^\dag(0)
                                         \!+\! c^{\dag}_{\mathbf{k}m \sigma}(0) c_{\mathbf{k}m \sigma}^{\vphantom{\dag}}(t) \Big\rangle\label{eq:G-ret-time}\\
  &G_R(\mathbf{k}, m, \omega) = \int_{-\infty}^\infty \!dt \;e^{i\omega t}\; G_R(\mathbf{k},m, t).\label{eq:G-ret-freq}
\end{align}
Here, $\Theta(t)=0$ for $t<0$ and
$1$ for $t>0$ is the step function; and  $\langle ... \rangle$ the grand canonical expectation value.
One can further define a self-energy
\begin{equation}
  \label{eq:dyson}
  \Sigma_R(\mathbf{k}, m,\omega) = \big[G_R^{(0)}(\mathbf{k}, m,\omega)\big]^{-1} - \big[G_R(\mathbf{k}, m,\omega)\big]^{-1} 
\end{equation}
as the difference between (inverse) non-interacting ($U=0$) Green's function $G_R^{(0)}(\mathbf{k}, m,\omega)$ 
and interacting ($U$) Green's function $G_R(\mathbf{k}, m,\omega)$, 
which contains all effects of the interaction\cite{Abrikosov1975a}. Here, and similarly in 
\begin{equation}
  \label{eq:g-nonint}
  G_R^{(0)}(\mathbf{k}, m, \omega) = \lim_{\alpha\rightarrow 0^+}\big[ \omega  + \mu + i\alpha - \epsilon_m(\mathbf{k}) \big]^{-1},
\end{equation}
the orbital-diagonal dispersion relation allows us to avoid matrix-inversions in the orbital indices; $\mu$ is the chemical potential.

In DMFT, which becomes exact in the limit of infinite dimensions \cite{Metzner1989}, 
the momentum dependence of the self-energy is neglected: $\Sigma_R(\mathbf{k}, m, \omega) \to \Sigma_R(m, \omega)$. 
Thus the one-particle Green's function of the Hubbard model in the DMFT approximation is
\begin{equation}
  \label{eq:g-ret-w-explicit}
  G_R(\mathbf{k}, m, \omega) = \big[ \omega +\mu - {\epsilon}_m(\mathbf{k}) - \Sigma_R(m,\omega)\big]^{-1},
\end{equation}
where the  $i\alpha$ of Eq.\ \eqref{eq:g-nonint} becomes obsolete since ${\rm Im}\Sigma_R(\omega)$ is negative.
For the actual calculation of this self-energy in DMFT, done  through a self-consistent solution of an Anderson impurity model, 
we refer the reader to Refs.~\onlinecite{Georges1992a,Georges1996,Held2007}.

Let us instead turn to our actual task, i.e.\ calculating scattering rates or
quasiparticle life times. For the following considerations we drop the orbital ($m$) dependence, 
as the Green's function and self-energy are anyhow diagonal in $m$ due to the assumed dispersion relation. 
If we linearize the real part of the self-energy and parameterize it through the quasiparticle weight $Z$, 
i.e., $\mathrm{Re}\Sigma_R(\omega)\approx\mathrm{Re}\Sigma_R(0)+[1-Z^{-1}] \omega$
we can approximate  Eq.~(\ref{eq:g-ret-w-explicit}) as
\begin{equation}
  \label{eq:G_QP}
  G_R(\mathbf{k}, \omega) \approx Z \big[ \omega - \tilde{\epsilon}(\mathbf{k}) -Z \mathrm{Im}\Sigma_R(\omega)\big]^{-1}, 
\end{equation}
where the Green's function has a quasiparticle pole at 
$\omega=\tilde{\epsilon}(\mathbf{k})=Z[{\epsilon}(\mathbf{k})+\mathrm{Re}\Sigma_R(0)-\mu]$,
with a Lorentzian broadening of full-width--half-maximum of
$-2Z \mathrm{Im}\Sigma_R(\tilde{\epsilon}(\mathbf{k}))$. 
That is, $\tilde{\epsilon}(\mathbf{k})$ is the quasiparticle energy and the broadening indicates that 
\begin{equation}
  \label{eq:scatrat-dmft-w}
  \frac{1}{\tau[\omega=\tilde{\epsilon}(\mathbf{k})]} = -2 Z \mathrm{Im} \Sigma_R(\omega=\tilde{\epsilon}(\mathbf{k})).
\end{equation}
is the inverse life time, also known as scattering rate.

Even more transparent is the role of the life time $\tau$ when we recapitulate 
the physical meaning of the time-dependent
retarded Green's function Eq.\ \eqref{eq:G-ret-time}. 
For the special case of zero temperature the system is in the ground state $|\mathrm{GS}\rangle$ and if the momentum $\mathbf{k}$  is not
occupied in the ground state,
Eq.\ \eqref{eq:G-ret-time} is reduced to
\begin{equation}
  G_R(\mathbf{k}, t) = -i\langle \mathrm{GS} | c^{\vphantom{\dag}}_\mathbf{k}(t) c^\dag_\mathbf{k}(0) | \mathrm{GS} \rangle \; .
\end{equation}
That is, at time $t=0$ a particle is added to the system which is thus in the state
$|\phi\rangle = c^\dag_\mathbf{k}(0) |\mathrm{GS}\rangle$.
Projecting this state onto its propagated version at time $t>0$
$\langle\phi(t)| = e^{i E_\text{GS} t} \langle \mathrm{GS} | c^{\vphantom{\dag}}_\mathbf{k}(t)$
yields the probability amplitude ($E_\text{GS}$ is the ground state energy) that this state still exists
after a time $t$ has elapsed \cite{Nolting2015}.
This motivates the interpretation of $|G_R(\mathbf{k}, t)|^2$
as the probability that a state created by  addition of a particle
at $t=0$ still exists at later time $t>0$. 

In Appendix \ref{app:gft}, we will show that this probability is approximately
\begin{equation}
  \label{eq:g-t}
  \big|G_R(\bold k, t)\big|^2 \propto e^{2Z\mathrm{Im}\Sigma_R(\tilde{\epsilon}(\mathbf{k}))\, t}\equiv e^{-t/\tau(\tilde{\epsilon}(\mathbf{k}))},
\end{equation}
which again leads to Eq.~(\ref{eq:scatrat-dmft-w}) for the  life time $\tau$.

Technically,  we calculate  the DMFT self-energy on Matsubara frequencies \cite{Matsubara1955}
by continuous-time quantum Monte Carlo \cite{Gull2011a} with symmetric improved estimators \cite{Kaufmann2019} using the w2dynamics program package \cite{Parragh2012,w2dynamics}.
The retarded self-energy at real (physical) frequencies is then obtained by the maximum entropy analytic continuation
\cite{Jarrell1996,Geffroy2019,kaufmannGithub}.

\subsection{Boltzmann scattering equation}
The key quantity of the BSE \cite{Boltzmann1872,Snoke2007,Ziman1960,Chambers1990} is the distribution function, whose dynamics is described
through the leading-order contributions of the  particle-particle interaction (for the models considered). 
In cases where elementary particles interact strongly, it is recommendable to rewrite the Hamiltonian 
in terms of weakly interacting quasiparticles so that the leading order
perturbation theory can be applied to the weaker effective quasiparticle interaction.

Here, we assume that a quasielectron description is possible and that these quasiparticles are characterized 
by a certain set of quantum numbers, namely the momentum $\bold k$, spin $\sigma$ or orbital-index $n$, 
and a corresponding quasiparticle dispersion relation $\tilde\epsilon_{n \sigma}(\bold k)$. 
Then the distribution function  $f_{n \sigma}(t,\bold k)$ corresponds to the expectation value of the occupation number operator
of a single-particle state $n_{\mathbf{k} n \sigma}$  at time $t$. 
In the following the spin will be absorbed into the band index for brevity.

The BSE in case of a spatially homogeneous system without external fields but with a
fermionic particle-particle  scattering reads \cite{Snoke2007,Ziman1960,Chambers1990}
\begin{equation} \label{eq:boltz1}
\begin{split}
&\frac{\mathrm d f_{n_0} (\bold k_0) }{\mathrm d t} = \frac{1}{2} \sum_{n_1 n_2 n_3} \int \mathrm d^d k_1 \mathrm d^d k_2 \mathrm d^d  k_3 \Big [ W_{n_0 \dots n_3} (\bold k_0\dots \bold k_3) \\
&\quad \times   \Big ( (1-f_{n_0}(\bold k_0)) (1-f_{n_1}(\bold k_1))f_{n_2}(\bold k_2) f_{n_3}(\bold k_3)\\
&\quad\quad \quad -f_{n_0}(\bold k_0) f_{n_1}(\bold k_1) (1- f_{n_2}(\bold k_2))(1- f_{n_3}(\bold k_3)) \Big ) \Big ]
\end{split}
\end{equation}
for a $d$-dimensional system. Here, $W_{n_0 \dots n_3} (\bold k_0\dots \bold k_3)$ is defined as 
\begin{equation}
\begin{split}
&W_{n_0 \dots n_3} (\bold k_0\dots \bold k_3) = w_{n_0\dots n_3} (\bold k_0 \dots \bold k_3)  \\ 
& \quad  \quad \times \delta(\tilde\epsilon_{n_0}(\bold k_0)+\tilde\epsilon_{n_1}(\bold k_1)-\tilde\epsilon_{n_2}(\bold k_2)-\tilde\epsilon_{n_3}(\bold k_3) ) \\
  & \quad  \quad \times \sum_{\bold G} \delta(\bold k_0 + \bold k_1 - \bold k_2 - \bold k_3 + \bold G) \; ;
  \label{eq:BSEscat}
\end{split} 
\end{equation}
and  the scattering amplitude $w_{n_0\dots n_3} (\bold k_0 \dots \bold k_3)$ 
can be calculated by perturbation theory (Fermi's Golden rule) and is $\sim U^2$
(explicit formulas follow in the context of the specific models below).
The two delta-distributions $\delta(\cdot)$ ensure momentum and energy conservation at the scattering event and 
the sum $\sum_{\bold G}$ runs over all reciprocal lattice vectors $\bold G$.
 
In thermal equilibrium, the distribution of electrons is given by the Fermi-Dirac distribution, 
$f_\textrm{FD}(\tilde\epsilon) = 1 / \big (1+\mathrm{exp}[\beta (\tilde\epsilon)] \big )$ 
with the inverse temperature $\beta=1/T$, and the chemical potential $\mu$ already absorbed in $\tilde\epsilon$.
The Fermi-Dirac distribution is a fixed point of the Boltzmann equation Eq.~\eqref{eq:boltz1} 
and therefore properly represents an equilibrium system. 

The scattering rate $1/\tau_n(\bold k)$ of a test-particle that is added in the state $(n,\bold k)$  in thermal equilibrium can be calculated within the Boltzmann framework as (for a derivation, see \cite{wais2020Preprint}):
\begin{equation}
\begin{split}
&\frac{1 }{\tau_{n_0}(\bold k_0)} = \frac{1}{2} \sum_{n_1 n_2 n_3} \int \mathrm d^d k_1 \mathrm d^d k_2 \mathrm d^d  k_3 \Big [ W_{n_0 \dots n_3} (\bold k_0\dots \bold k_3) \\
&  \times   \Big ( (1-f_\textrm{FD}(\tilde\epsilon_{n_1}(\bold k_1)) )f_\textrm{FD}(\tilde\epsilon_{n_2}(\bold k_2)) f_\textrm{FD}(\tilde\epsilon_{n_3}(\bold k_3))\\
&  + f_\textrm{FD}(\tilde\epsilon_{n_1}(\bold k_1)) (1- f_\textrm{FD}(\tilde\epsilon_{n_2}(\bold k_2)))(1- f_\textrm{FD}(\tilde\epsilon_{n_3}(\bold k_3))) \Big )\Big ].
\end{split} \label{eq:boltzwithk}
\end{equation}
The  calculation of the scattering rate above is done numerically with the method presented in Ref.~\onlinecite{wais2020Preprint}.
Notice that DMFT scattering rates are only energy (and orbital) dependent. 
In the BSE we can, on the other hand, add a quasiparticle at every momentum $\bold k$
which then necessarily has the quasiparticle energy $\tilde\epsilon_{n}(\bold k)$. 
When we later plot the BSE scattering rates as a function of energy,
there will be different $1/\tau_n(\bold k)$'s at the same energy $\tilde\epsilon$.
Note that the many-body life time broadening discussed above also allows us to add particles away from
$\tilde\epsilon_{n}(\bold k)$ in DMFT, albeit the spectral density of such states is strongly suppressed if the broadening is weak.

\subsection{BSE without momentum conservation}
Prospective differences between the BSE and DMFT may emerge because  of
(i) strong coupling effects beyond the perturbative treatment of the scattering  in the BSE rate equation and
(ii) neglecting the momentum dependence in DMFT.
The latter not only reflects in the momentum-independent DMFT self-energy 
but also in disregarding the momentum conservation at scattering events in DMFT. 
That is, the DMFT self-energy is calculated from Feynman diagrams to all order in $U$
but with the interaction only on an impurity which per construction breaks momentum conservation.

We can apply the same approximation also to Boltzmann scattering.
That is, we remove in Eq.~(\ref{eq:BSEscat}) the momentum conserving delta-distributions 
$\sum_{\bold G} \delta(\bold k_0 + \bold k_1 - \bold k_2 - \bold k_3) \to \frac{1}{V_{BZ}}$, 
where $V_{BZ}$ is the volume of the first Brillouin-zone, as was proposed in Ref.\ \onlinecite{wais2018}. 
Eq.~(\ref{eq:boltzwithk}) can then be simplified to a purely energy-dependent scattering rate 
$1/\tau_n(\epsilon)$ that is calculated as \cite{wais2018,wais2020Preprint}
\begin{equation}
\begin{split}
&\frac{1}{\tau_{n_0}(\epsilon_0)} = \frac{1}{2} \sum_{n_1 n_2 n_3} \int \mathrm d \epsilon_1 \mathrm d \epsilon_2 \mathrm d \epsilon_3 \Big [ \tilde w_{n_0 \dots n_3} (\epsilon_0\dots \epsilon_3)\\
& \quad \times \delta(\epsilon_0 + \epsilon_1 - \epsilon_2 - \epsilon_3) A_0^{n_1}(\epsilon_1) A_0^{n_2}(\epsilon_2) A_0^{n_3}(\epsilon_3)\\
&\quad \quad \quad \times \Big ( (1-f_\textrm{FD}(\epsilon_{1}) )f_\textrm{FD}(\epsilon_{2}) f_\textrm{FD}(\epsilon_{3})\\
& \quad\quad \quad + f_\textrm{FD}(\epsilon_{1}) (1- f_\textrm{FD}(\epsilon_{2}))(1- f_\textrm{FD}(\epsilon_{3})) \Big )\Big ] ,
\end{split} \label{eq:boltznok}
\end{equation}
where $A_0^n(\epsilon)$ is the normalized DOS of band $n$ 
and $\tilde w_{n_0 \dots n_3} (\epsilon_0\dots \epsilon_3)$ is the thus modified scattering amplitude
that depends on the energies only. 

Notice that in Eq.~(\ref{eq:boltznok}) we have explicitly used the fact that the interaction is itself momentum-independent 
(which is the case for the purely local interaction  in the Hubbard model).
In the general case Eq.~(\ref{eq:boltznok}) cannot be derived, 
but it can be constructed as an approximation~\cite{Ono_2018,Ono_2020}.
In the following we will refer to Eq.~\eqref{eq:boltznok} as Boltzmann without momentum conservation (BSE without $\bold k$).

Note that the structure of Eq.~\eqref{eq:boltznok} is way simpler than Eq.~(\ref{eq:boltzwithk}):
it can be computed by inverting analytically the energy-conserving delta distribution in Eq.~\eqref{eq:boltznok}
and then using standard numerical integration techniques.

\section{One-band Hubbard model at weak coupling} \label{chap:weak1band}
\begin{figure*}
 \includegraphics[width=15cm]{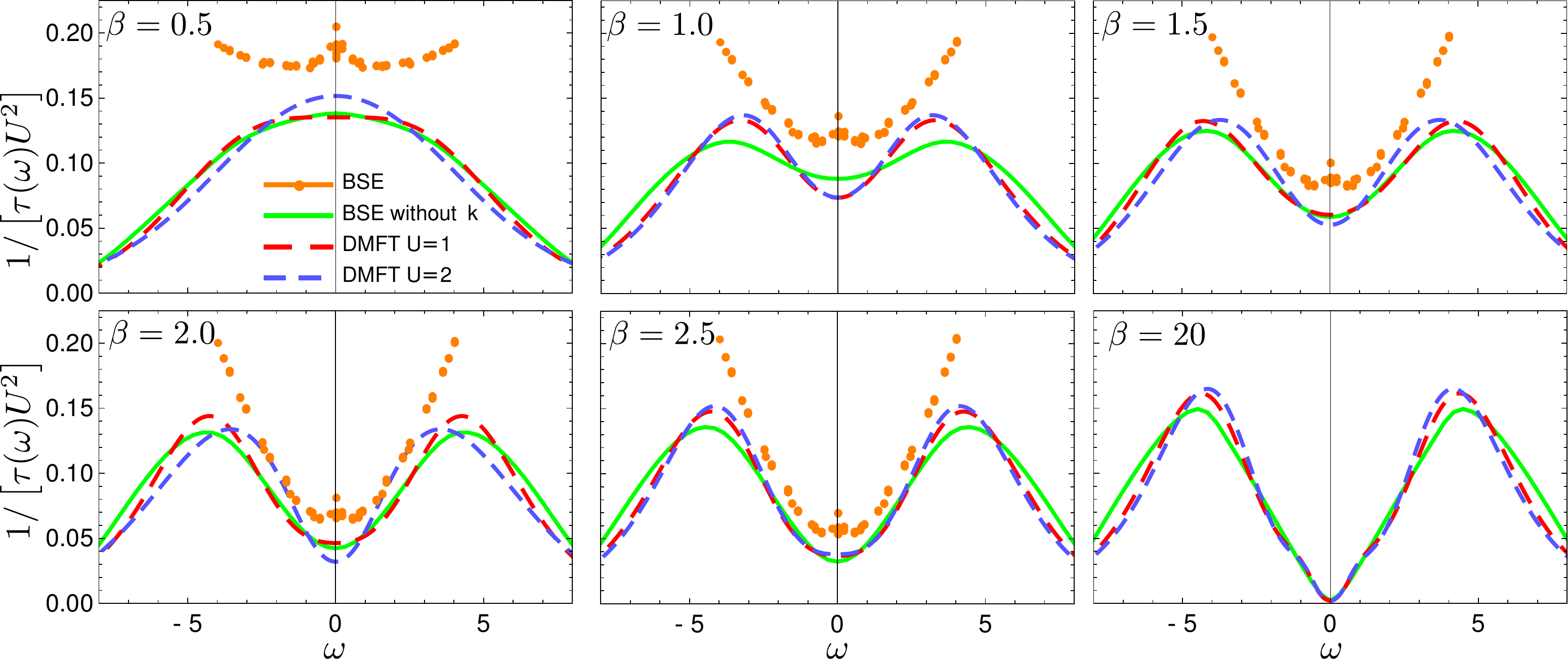}
 \caption{Scattering rates $1/\tau$ normalized by the interaction squared ($U^2$) for the two-dimensional Hubbard model 
  at half-filling calculated by DMFT and BSE with and without $\mathbf k$ conservation.
  The case $\beta=20$ could not be calculated with full Boltzmann due to computational limitations.
  The scattering rates shown are the same for both spins in the paramagnetic phase.}\label{fig:1band}
\end{figure*}

As a first comparison, we  discuss the case of the prototypical 
one-band Hubbard model in two dimensions at half-filling. 
Depending on the strength of the local interaction $U$ and the temperature $T=1/\beta$,
such a system is predicted by DMFT to be either metallic or Mott-insulating. 

For the weak coupling case we may employ Boltzmann theory with the dispersion relation of the non-interacting Hamiltonian, which is
\begin{equation}
\begin{split}
  &\epsilon(\mathbf{k}) = -2 t [\cos(k_x) + \cos(k_y)] \label{eq:disprel}
\end{split} 
\end{equation} 
for  ${\bold k}=(k_x,k_y) \in [-\pi,\pi) \otimes [-\pi,\pi)$ (lattice constant $a\equiv 1$; unit-cell volume $V_\text{UC}=a^2=1$)
and the corresponding DOS \cite{Katanin2012}
\begin{equation}
  A_0(\omega) =  \int_\text{BZ} \!\! \frac{ d^2{k}}{V_\text{BZ}} \ \,\delta(\omega \!-\! \epsilon(\bold k))
  = \frac{1}{2\pi^2t} \mathrm{K}\Bigg(\!\!\sqrt{1 - \left(\frac{\omega}{4t}\right)^2}\Bigg)\textrm{} \label{eq:dosNonInt1B}
\end{equation}
where $\mathrm{K}(\ldots)$ is the complete elliptic integral of first kind.
As hopping parameter and unit of energy we choose $t\equiv 1$ in the following. 
The scattering amplitude for this system can be calculated in perturbation theory as 
\begin{equation}
  w (\bold k_0 \dots \bold k_3) = \frac{2 \pi}{{V_{BZ}}^2} U^2 \delta_{\sigma_0 \bar \sigma_1} \delta_{\sigma_2 \bar \sigma_3}
\end{equation}
with the short-hand notation $\bar \sigma_i \equiv - \sigma _i$ 
for the BSE scattering rate Eq.~\eqref{eq:boltzwithk},
and
\begin{equation}
  \tilde w (\epsilon_0 \dots \epsilon_3) = 2 \pi U^2 \delta_{\sigma_0 \bar \sigma_1} \delta_{\sigma_2 \bar \sigma_3 }
\end{equation}
for the case of BSE without $\mathbf{k}$ in Eq.~\eqref{eq:boltznok}, cf.~Ref.\ \onlinecite{wais2018}.

In Fig.~\ref{fig:1band} we show the calculated scattering rates for different temperatures comparing DMFT and the BSE  with and without momentum conservation. The quasi-particle renormalization is $Z\approx 1$ for these values of the interaction.
In order to compare the structure of the scattering rates for different interaction strengths, we divide the scattering rate by $U^2$. 
The Boltzmann scattering rates then become completely independent of $U$.
In contrast, the DMFT scattering rates depend on $U$ in a non-trivial fashion
(Fig.~\ref{fig:1band} shows $U=1$ and $U=2$) since it is a non-perturbative approach. Nonetheless in the limit  $U\rightarrow 0$, the DMFT normalized scattering rates must be $U$-independent.

Comparing the DMFT scattering rates for both interaction strengths,  one notices that the thus normalized scattering rates lie almost on top of each other for the inverse temperatures 
$\beta = 1.0$, $\beta = 2.5$ and $\beta = 20$, while they slightly deviate 
for $\beta = 0.5$, $\beta = 1.5$, $\beta = 2.0$. Since there is a rather large uncertainty 
from the maximum entropy analytical continuation and the deviation is
not systematic, we can  conclude that the differences in the normalized
DMFT scattering rates at $U=1$ and $U=2$ are within the error bars.

\begin{figure}
  \includegraphics[width=6.5cm]{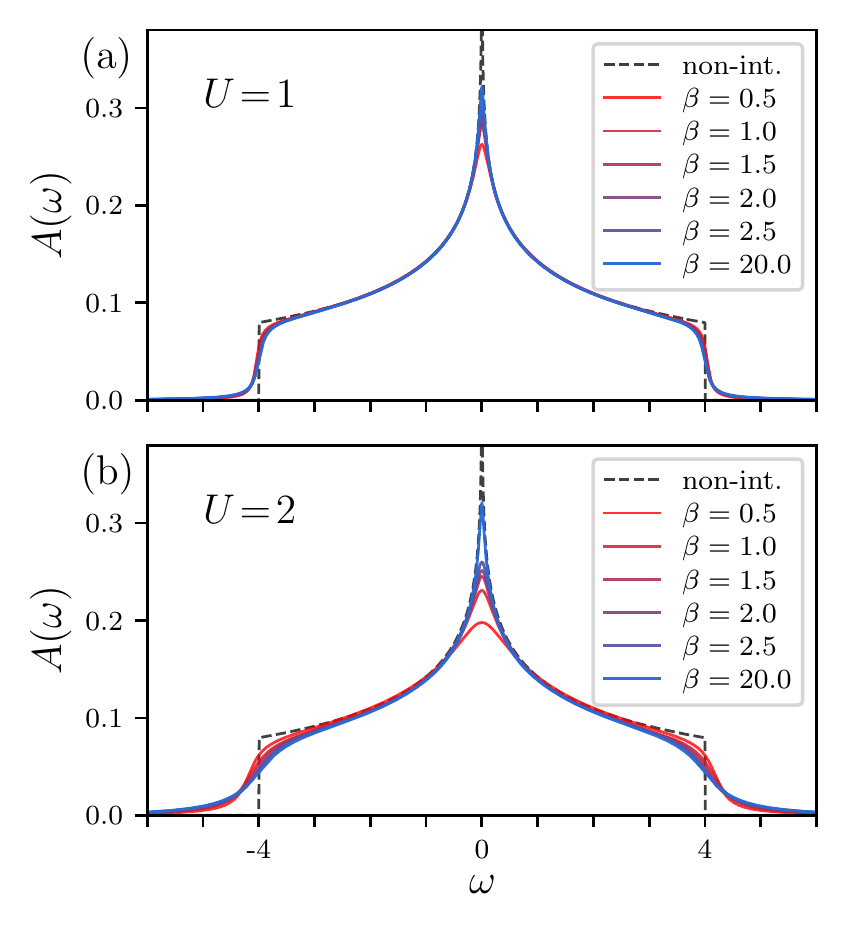}
  \caption{\label{fig:1band_U1_U2_specdens} DMFT spectral densities for (a) $U=1$ 
    and (b) $U=2$ for different temperatures}
\end{figure}

The  scattering rates calculated by the BSE without $\bold k$ are in very good agreement with the DMFT data
for all inverse temperatures except for $\beta = 1.0$. 
Again, this deviation may well originate from the uncertainties of the analytic continuation.
In any case, the good agreement of the scattering rates from BSE without $k$ and DMFT along with the $\sim U^2$ scaling of the DMFT results, clearly show that even at $U=2t$ we are still in the perturbative regime. As we show in Appendix~\ref{sec:IPT}, to second order in $U$ the scattering rates as calculated in DMFT and  BSE without $\bold k$ are indeed identical.

Note however that the spectral density\footnote{
  The DMFT spectral densities were calculated as $-\mathrm{Im}G_R(\omega)/\pi$ with
  \begin{equation}
    \label{eq:specdens-nice}
    G_R(\omega) = \int_{-\infty}^{\infty} dx \frac{A_0(x)}{\omega-x-\Sigma_R(\omega)}
  \end{equation}
  after analytical continuation of the self-energy.
  This allows for resolving features like sharp peaks in the spectral density
  that would be smeared out by direct analytic continuation of the local
  Green's function in Matsubara frequencies.}
in Fig.~\ref{fig:1band_U1_U2_specdens} is already significantly smeared,
especially at the band edges and the Van-Hove singularity, because of the stronger interaction. 
This smearing is a direct consequence of the scattering rate in Fig.~\ref{fig:1band};
and through the DMFT self-consistency it will in turn affect the scattering rates, 
but only in higher order in $U$ (when self-consistently calculating the spectral function 
as indicated in  Appendix~\ref{sec:IPT}). 
Possibly this explains why the BSE without $\bold k$ in Fig.~\ref{fig:1band} has a lower scattering rate at the band edge $\omega=\pm4$
and a larger one for larger $|\omega|$, albeit we cannot exclude this to be an artifact of the analytical continuation. 

Both DMFT and BSE without $\bold k$
show a two-peak structure in the scattering rates with the peak positions roughly at the band-edges.
The width of these peaks increases with temperature. 
At the highest temperature ($\beta = 0.5$) there is only one peak visible which actually consists of the two peaks that are strongly overlapping.
In Appendix~\ref{chap:appConvolution}, we show that the position of the two peaks can be approximately calculated from the first moment of the particle- and hole-density. The width and height of the peaks can be calculated when the zeroth and second moment of the particle-density is taken into account in addition to the first moment.

After establishing a good agreement between DMFT and BSE without $\mathbf k$ at weak coupling, we next turn to the full BSE with momentum conservation.
The thus calculated BSE scattering rates (dots in Fig.~\ref{fig:1band}) deviate from the rates obtained with the other methods. First of all, as already mentioned, we highlight that several values, corresponding to different momenta, are present for each energy.  Fig.~\ref{fig:1band} shows a particularly strong spread at the Fermi level ($\omega=0$).
Furthermore, in contrast to BSE without $\bold k$ and DMFT, there are no scattering rates  outside the non-interacting bandwidth ($|\omega|>4$)  any longer, as there is no momentum that has such an energy. In DMFT due to the aforementioned smearing of the band-edges there are such states, and in  BSE without $\bold k$ we can at least calculate the scattering rate a state at such an energy would have.

Another difference is that 
the BSE scattering rates are generally higher than DMFT or BSE without $\mathbf k$, especially at the band edge ($|\omega|\lesssim 4$ and at higher temperatures also around the Fermi level ($\omega=0$).
As DMFT and BSE without $\bold k$ agree with each other, we can safely conclude that this difference  originate from neglecting the momentum conservation of the scattering vertex. One can also smoothly interpolate between the results for the BSE with and without $\mathbf k$, by replacing the momentum conserving $\delta$-function by a Gaussian and increasing its width (not shown here). 
The reason for these discrepancies is that the momentum averaged scattering amplitude does not take into account that there is  e.g.\ a particularly strong scattering among  momenta at the Hove singularities  $(\pm \pi,0)$ and $(0,\pm \pi)$.
At low temperatures this scattering even leads to the formation of a pseudogap \cite{Vilk1997,Norman1998,Timusk_1999,Keimer2015,RevModPhys.78.17,Sordi2012,PhysRevLett.114.236402,PhysRevX.8.021048}
which requires a beyond DMFT description \cite{Sadovskii2005,Zhang2007,Katanin2009,Gull2013,Schaefer2015-2,RMPVertex}. A precursor thereof is visible here as the strong-momentum dependence of the scattering rate on the Fermi surface.

\section{Two-orbital band insulator} \label{weakCoupling2Band}
\begin{figure}
 \includegraphics[width=6.5cm]{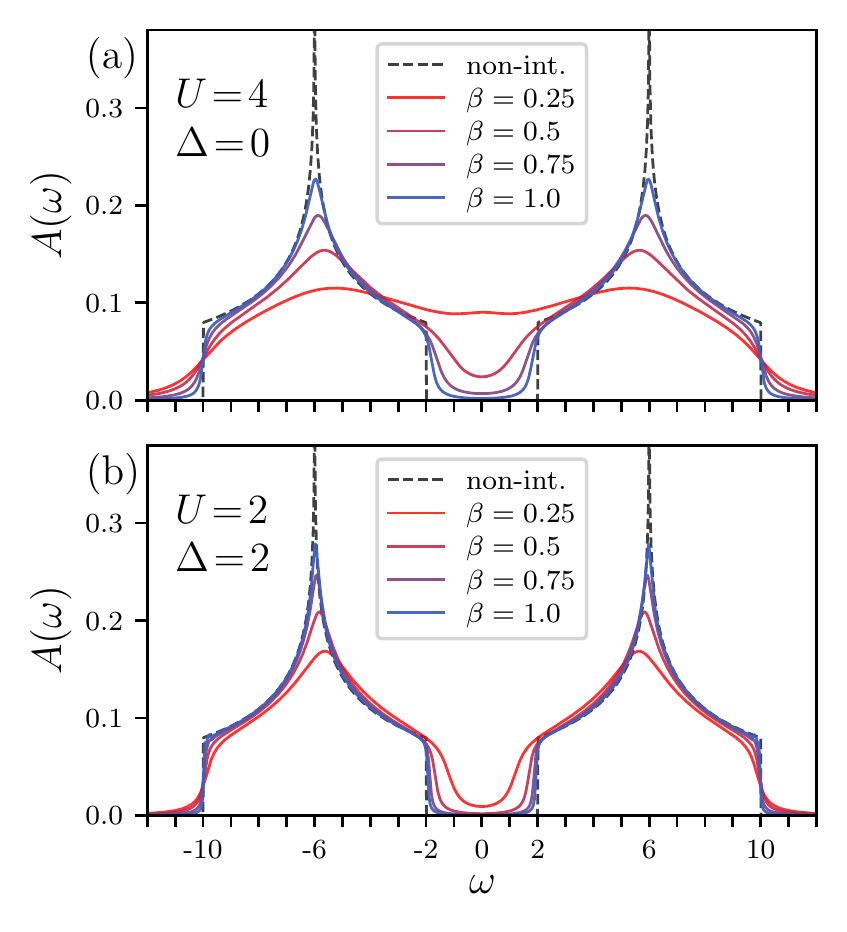}
 \caption{Spectral-densities for different temperatures for the case (a) $U=4$, $\Delta = 0$ and (b) $U=2$, $\Delta = 2$. For both cases, the effective band gap is $\Delta_\textrm{eff.} \approx 4$.}\label{fig:2bandSpecDens}
\end{figure}
In this section, we address the case of a band insulator in the weak to intermediate coupling regime. 
We consider a two-dimensional Hubbard-type model with two orbitals ($A$ and $B$) at half-filling, 
i.e., $n=2$ electrons per site in the two orbitals. This corresponds to $\mu = 0$ for our dispersion relation below.
For simplicity, we assume that electrons may only hop to neighboring orbitals of the same type
and that  the hopping amplitude has the same absolute size but opposite  sign
for both orbitals ($t_A=-1$, $t_B=1=t$).
Further, we add a local one-particle energy $\mp( \Delta /2 + 4t)$ for  orbital $A$ and $B$, respectively.
This results in a band gap of size $\Delta$ in the non-interacting DOS, with the top of the valence($A$)-band  and the bottom of the 
conduction($B$)-band both at the $\Gamma$ point.
The interaction $U$ is local and the same within and between both orbitals 
such that the interaction term of the Hubbard model acquires
the simple form of Eq.\ \eqref{eq:hubbard-hamiltonian}.

We now discuss two different systems, one with $U=4$ and one-particle gap $\Delta = 0$ and one with $U=2$ and $\Delta = 2$. 
Due to the constant Hartree term in the self-energy, 
the effective gap in the interacting system  is essentially the same $\Delta_\textrm{eff.} \approx U + \Delta= 4$ for both setups.
This is because at sufficiently low  temperatures, orbital $A$ is almost completely filled with two electrons per site and orbital B is empty. Hence an electron in orbital $B$ perceives a Hartree energy $2U$ (interacts with both $A$ electrons); an electron in orbital $A$ instead has a Hartree energy $1U$ (as it only interacts with the electron of opposite spin in orbital $A$). The difference enlarges the bandgap to $\Delta_\textrm{eff.} = U + \Delta$.

The spectral densities for both cases are displayed in Fig.~\ref{fig:2bandSpecDens} and follow the above reasoning. 
At higher temperatures, we however induce holes in the valence and electrons in the conduction band.
The difference in occupation is reduced, the bandgap hence smaller.
For the highest temperature ($\beta = 0.25$), the gap disappears completely for the case $U=4$, $\Delta = 0$. 
The non-interacting DOS in  Fig.~\ref{fig:2bandSpecDens} is  constructed with the above enhanced effective band gap
$\Delta_\textrm{eff.}$ instead of $\Delta$.

As this describes the DMFT spectrum at low temperatures reasonably well, we employ for the BSE the corresponding effective  bandstructure 
\begin{align}
\epsilon_A(\bold k) =& -\epsilon(\bold k) - \left ( \frac{\Delta_\textrm{eff.}}{2} + 4t \right ) ,\\
\epsilon_B(\bold k) =& \epsilon(\bold k) +\left ( \frac{\Delta_\textrm{eff.}}{2} + 4t \right ) ,
\end{align}
where $\epsilon(\bold k)$ is defined by Eq.~\eqref{eq:disprel}. 
The corresponding DOS of the 
non-interacting system corrected by the Hartree shift  is used for the BSE without $\mathbf k$ and given by
\begin{align}
A_0^A(\omega) = & A_0 \left ( \omega + \left ( \frac{\Delta_\textrm{eff.}}{2} + 4t \right ) \right ) ,\label{eq:A0A}\\
A_0^B(\omega) = & A_0^A \left (-\omega \right ) \label{eq:A0B}
\end{align}
with $A_0(\omega)$ defined in Eq.~\eqref{eq:dosNonInt1B}. 
Due to particle-hole symmetry and the simple form of the interaction,
the BSE calculation can be simplified as outlined in Appendix~\ref{chap:simp2band}.

\begin{figure*}
 \includegraphics[width=12cm]{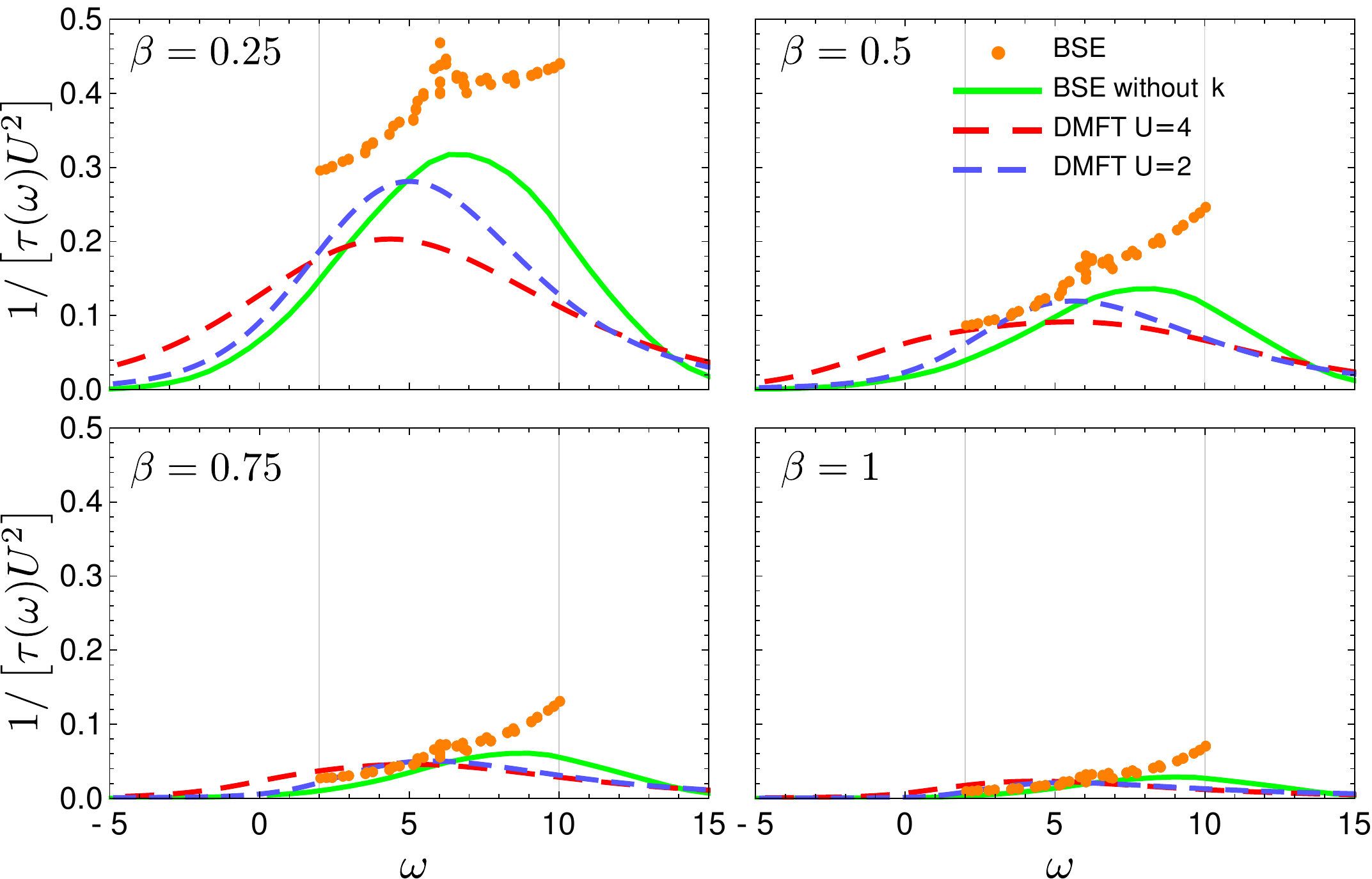}
 \caption{Scattering rates normalized by the squared interaction for an electron in the upper band of a two-orbital band insulator as calculated with DMFT, BSE with and without $\mathbf k$. Two different sets of parameters are used: $U=4$, $\Delta = 0$ and $U=2$, $\Delta = 2$. The gray, vertical lines indicate the band edges of the non-interacting system.}\label{fig:2band}
\end{figure*}

Fig.~\ref{fig:2band} shows the scattering rate of the two-band system. 
The BSE with momentum conservation shows a seemingly parabolic increase starting with a sizable value 
at the lower band edge ($\omega=2$). Superimposed on this trend is an enhanced scattering rate in the middle of the band
at $\omega=6$ with a strong momentum spread of the scattering rate. 
This is akin to the behavior at the Fermi level for the weakly correlated one-band Hubbard model in
Fig.~\ref{fig:1band} and can again be attributed to the van Hove singularity.

Similar as for the one-band case, the scattering rates in BSE without $\bold k$ are slightly smaller than in the BSE with  $\bold k$ conservation and already decay toward the upper band edge ($\omega=10$). They closely 
resemble the DMFT values for the $U=2$ case; 
only the peak of the scatterings is shifted to slightly higher energies than in DMFT.
There are  larger differences to the  DMFT data at the intermediate coupling $U=4$,
which have systematically higher scattering rates at low energies.
This is because stronger smearing of the spectral density at $U=4$ leads to a smaller effective gap
and some in-gap spectral weight, see Fig.~\ref{fig:2band}.
This, in turn, leads to more thermal excitations and therefore more scatterings.
These effects can be included in
 BSE without $\bold k$ if we use the interacting spectral density $A^n(\omega)$
instead of the non-interacting one $A_0^n(\omega)$, which leads to a good agreement with the DMFT results,
see Appendix~\ref{chap:twoBandInteracting}.

Eye catching is the strong suppression of the scattering rate upon decreasing temperature.
The reason for this is the dramatic reduction of the number of 
thermally excited carriers which are needed to act as scattering partners.
Note that with a density-density Coulomb interaction, the electron in the conduction($B$)-band either needs
(i) another $B$-electron to scatter with [the final state  being again two $B$-electrons], or 
(ii) a hole in the valence($A$)-band into which an $A$-electron can scatter 
[final and initial state being one $A$- and one $B$-electron]. 
Both $B$-electron and $A$-hole scattering partners however require
thermally excited carriers that are absent at low temperatures.

For the Mott insulator discussed in the next section, 
the scattering rates are much higher because of impact ionization processes. 
Here, an electron in the upper Hubbard band excites an additional electron-hole pair across the gap. 
In the band insulator impact ionization corresponds to a process
${c^{\dag}_{i B \sigma}c^{\vphantom{\dag}}_{i A \sigma}c^{\dag}_{i B \bar\sigma }c^{\vphantom{\dag}}_{i B \bar\sigma}}$
which is not possible in lowest order perturbation theory in the density-density interaction, 
nor are Auger processes
$c^{\dag}_{i B \sigma}c^{\vphantom{\dag}}_{i A \sigma}c^{\dag}_{i A \bar\sigma }c^{\vphantom{\dag}}_{i A \bar\sigma }$.
In the one-band Mott insulator, the two Hubbard bands have the same orbital index and such processes hence dominate the scattering rate 
if $\omega$ is sufficiently large to allow impact ionization~\cite{Werner2014,Sorantin2018,wais2018,Maislinger2020,Kauch2020a}.

Even if we generalize the Coulomb interaction to the widely employed Kanamori form \cite{Kanamori63}
with spin-flip and pair-hopping terms, we still need a thermally excited second electron or hole for scattering. 
Only, the full Slater  interaction \cite{Slater,Griffith} also contains interaction terms that directly mediate impact ionization.
These interaction terms are however small or even vanish, which is the reason why they are often disregarded in the first place. Consider e.g.\ a material with cubic symmetry and the orbitals $A=d_{xy}$ and  $B=d_{xz}$.
Then interaction terms such as
$U_{BAAA}c^{\dag}_{i B \sigma}c^{\vphantom{\dag}}_{i A \sigma}c^{\dag}_{i A \bar\sigma }c^{\vphantom{\dag}}_{i A \bar\sigma }$ vanish because the integral to calculate the matrix element $U_{BAAA}$ is odd under the transformaton $z\rightarrow -z$;  for a furthergoing discussion, see e.g.~\cite{Ribic2014,Buenemann2017}.
A more viable route to enhance the scattering rate through impact or Auger processes in a band insulator
is if the bands strongly hybridize so that the conduction and valence bands are admixtures of the $A$ and $B$ orbitals.

It is interesting to note that the scattering rate preserves its two-band like structure 
even in the case  $U=4$, $\beta = 0.25$ when the spectral density does not show a gap any longer.
The reason for this is again that the density-density interaction does not allow for impact excitation and Auger emission,
which are very gap-size sensitive. Instead the scattering processes
induced by the density-density interaction  are  agnostic about the gap-size per-se.
The additional $B$ electron still needs another $B$ (or $A$) electron to scatter with,
and two empty final $B$ states (or an empty  $A$ and an empty $B$ state).
The scattering process does not need to overcome the size of the gap,  in contrast to impact ionization.

\section{Strong coupling: Mott-insulator}\label{sec:mott}

\begin{figure}
 \includegraphics[width=7.4cm]{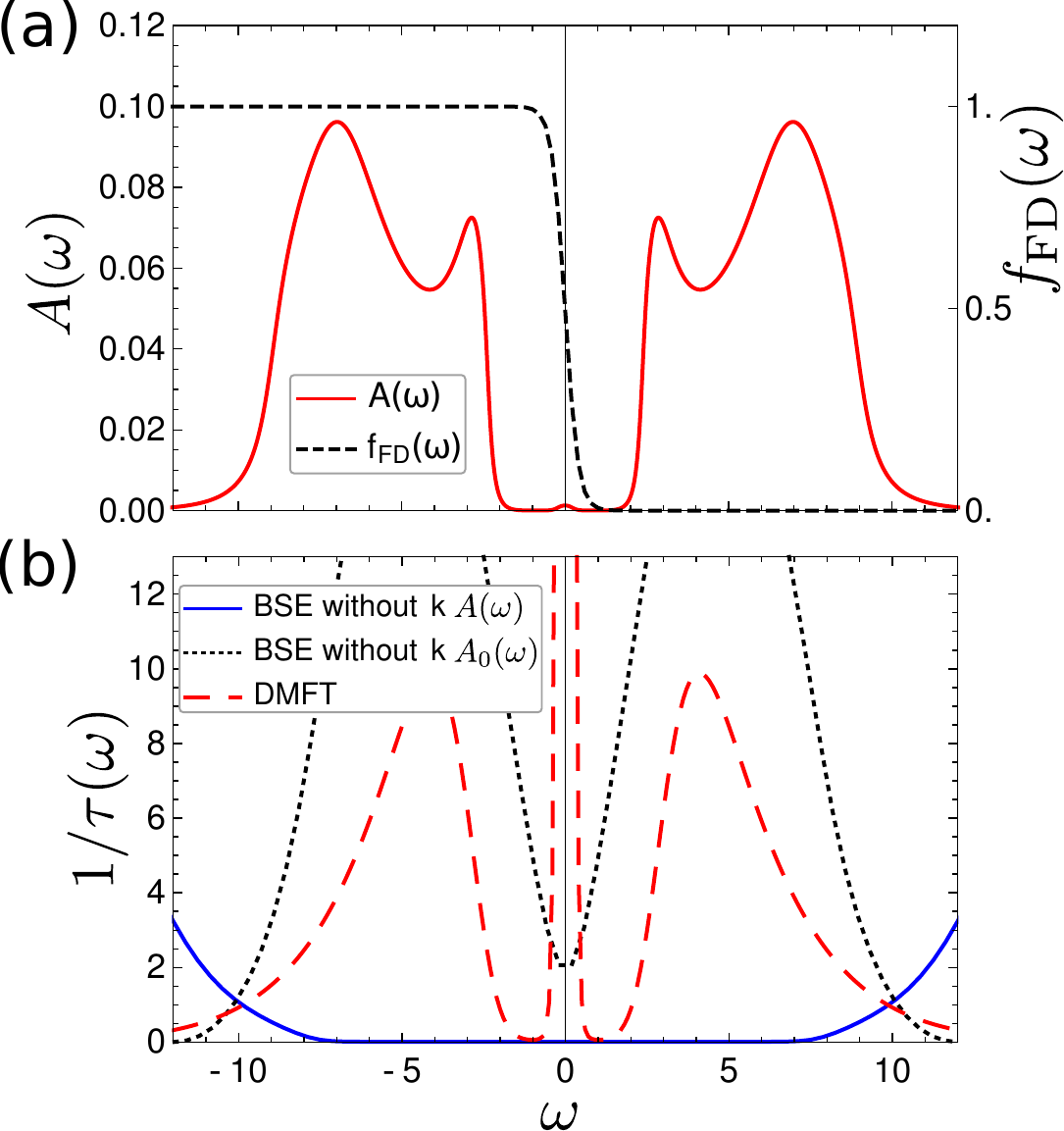}
 \caption{(a) Spectral density as obtained by DMFT and Fermi-Dirac distribution for the case $U=12$ and $\beta = 5$. (b) Scattering rates as obtained from DMFT, and BSE without $\mathbf k$  using either the non-interacting density of states (BSE without k $A_0(\omega)$) or the interacting DMFT spectral density shown (BSE without k $A(\omega)$).}\label{fig:mott_u12_b5}
\end{figure}

Finally, we compare the approaches introduced above in the strong coupling regime of the single-orbital Hubbard model.
Since the BSE is a perturbative treatment in the interaction, this is certainly the most problematic case for the BSE.
For sufficiently large interaction, the non-interacting DOS splits into two,
the upper and lower Hubbard band, see  Fig.~\ref{fig:mott_u12_b5}  (top).
We have a Mott insulator, one of the cornerstones of strongly correlated electron systems \cite{Gebhard1997}.

If we use the BSE with the non-interacting DOS, this dramatic reshuffling of the DOS is not incorporated.
The scattering rate is still the very same with a two peak structure as for weak coupling---just 
with the prefactor rescaled by $U^2$, see black-dotted line in Fig.~\ref{fig:mott_u12_b5}. 
This kind of description assumes that we have a metal with states at low energies. 
It is not an appropriate description of a Mott insulator.

This problem can be mitigated if we consider better suited quasiparticles
instead  of the non-interacting ones. This is in general not trivial,
and not always can proper quasiparticles with a long life time and weak interaction be identified.
They might not even exist.  Taking the electronic DMFT excitations of the Hubbard bands as our quasiparticles in the BSE
without $\bold k$, we have
to replace the non-interacting DOS  $A_0(\omega)$ by the
interacting spectral density  $A(\omega)$ of Fig.~\ref{fig:mott_u12_b5} (top) 
in Eq.~\eqref{eq:boltznok}. Even if we have no well defined quasiparticles 
such a quantum Boltzmann description is possible \cite{wais2018} if we have a separation of time scales,
and the average-time (distribution function) dynamics  is slower than the relative-time dynamics.
As was shown in  \cite{wais2018} the thus modified BSE without $\mathbf k$
provides a good description of the DMFT impact ionization processes and redistribution
of spectral weight in non-equilibrium\footnote{The calculation of scattering in this paper
is still possible within equilibrium DMFT theory; whereas the non-equilibrium processes of
\cite{wais2018} required the non-equilibrium DMFT~\cite{zlatic2006,Aoki2014}.}

Here, we instead study in Fig.~\ref{fig:mott_u12_b5} (bottom, blue line) the one-particle scattering rates
in the BSE without $\mathbf k$ and interacting $A(\omega)$:
The Mott insulator is described as two split quasiparticle bands with the
gap $\sim 4$ being much larger than temperature $T=1/5$.  
Hence, if we add an extra electron in the upper Hubbard quasiparticle band
it has no partners to scatter in BSE, the scattering rate is zero similar to the suppression of the scattering rate
in the band insulator. However, if the added electron has an excess energy
[$\omega - \omega_{LBE}$ relative to the lower band edge of the upper Hubbard band $\omega_{LBE}\gtrsim 2$ 
in Fig.~\ref{fig:mott_u12_b5}] which is larger than the Mott gap [$\Delta_{\rm Mott}\gtrsim 4$], 
i.e., $\omega\gtrsim 6$, impact ionization processes with an electron-hole excitation across the gap become possible.
The phase space of such scattering processes increase quadratically with $\omega- \omega_{LBE}-\Delta_{\rm Mott}$ for a box shaped DOS.
This explains the BSE without $\mathbf k$ scattering rate
in Fig.~\ref{fig:mott_u12_b5}, which as already mentioned well describes impact ionization processes,
including the change of the double occupation and redistribution of spectral weight with time in non-equilibrium~\cite{wais2018}.

Let us now turn to the DMFT scattering rate as extracted from the self-energy
and shown in Fig.~\ref{fig:mott_u12_b5} (bottom, red-dashed line) \footnote{As we 
do not have a linear quasiparticle renormalization in the self-energy, 
we plot $1/\tau(\omega)=2{\rm Im} \Sigma (\omega)$; $Z=1$ in Eq.~(\ref{eq:g-t}).}. 
The by far dominating feature (cut-off by the finite  $y$-axis scale) is at $\omega=0$ 
where $\Sigma(\omega)= (U^2/4) \; 1/(\omega+i\alpha)$ in the large $U$ limit of the Mott insulator 
with a Lorentzian broadening $\alpha \sim \pi T$. 
This pole is responsible for the splitting of the DOS into two Mott bands and yields the $\delta$-like peak
in ${\rm Im} \Sigma$ at $\omega=0$. As a matter of course we cannot expect this feature
to be described in the BSE without $\mathbf k$. It is also not necessary as $\omega=0$
is in the middle of the Mott gap where there are essentially no states---essentially
since at low temperature the aforementioned finite broadening leads to a very small spectral weight.
This filling of the Mott gap with temperature \cite{Mo2004} is a feature distinct from
a band insulator. These in-gap states have an extremely short life time.

\begin{figure}
  \centering
 \includegraphics[width=8.6cm]{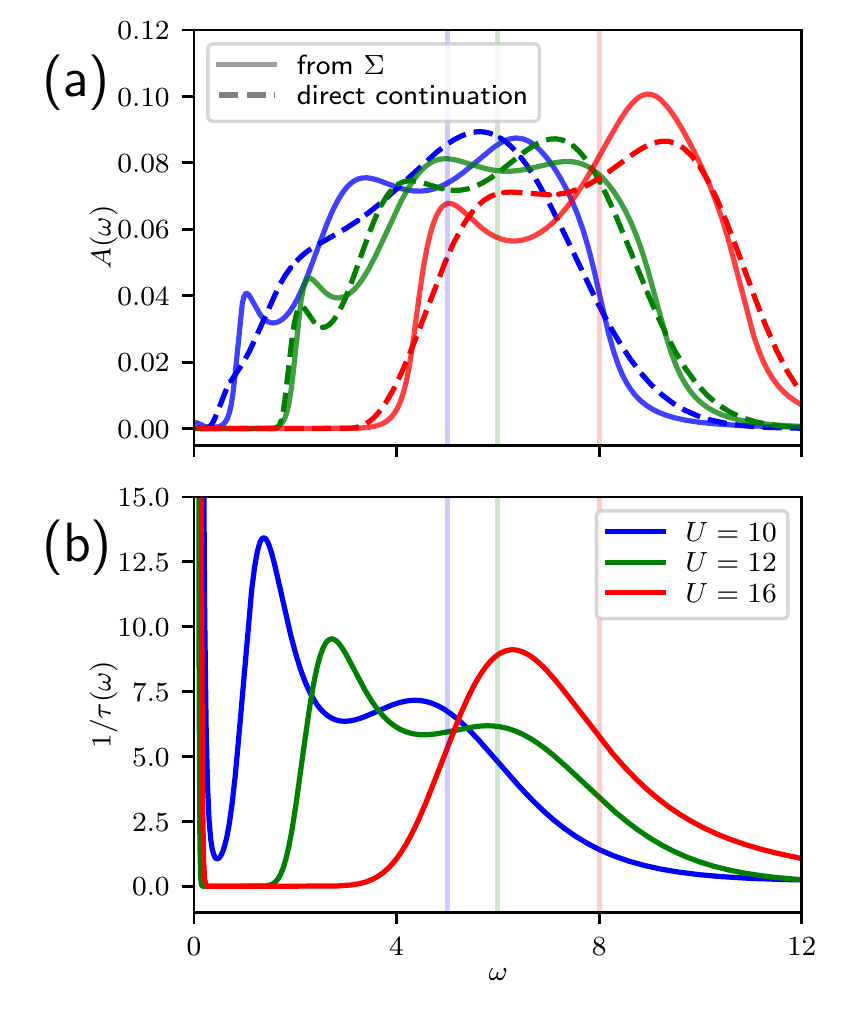}
 \caption{(a) Spectral density as obtained by DMFT for $\beta=10$ and different interaction strengths $U=10$, 12, 16. 
  The solid line $A(\omega)$'s are calculated from the analytically continued $\Sigma(\omega)$; 
  the dashed lines are directly analytically continued from the Matsubara Green's function. 
  (b) DMFT scattering rates for the same parameters as in (a).}\label{fig:mott_u}
\end{figure}
  
Let us now turn to the more relevant DMFT scattering rate within the Hubbard bands.
These are orders of magnitude larger in DMFT than those from the BSE
without $\mathbf k$ and with interacting $A(\omega)$.
Also their shape is completely different: There is no suppression at the lower edge of the upper Hubbard bands which,
as argued above, was the case if the scattering is due to  impact ionization
requiring a threshold energy; neither are the DMFT scattering rates flat or follow the
shape of the upper Hubbard band. Instead the scattering rates are strongest around $\omega\sim 4$ 
close to the lower band edge, and are dramatically reduced for larger $\omega$.
Similar as the pole at $\omega=0$, the maximum at  $\omega\sim 4$ leads to a suppression of the spectral weight.
Fig.~\ref{fig:mott_u12_b5} (top) where we have calculated $A(\omega)$ from the analytically continued $\Sigma(\omega)$
even shows a two peak structure in the upper Hubbard band. Such a two peak structure was previously observed
on the metallic side of the Mott transition, 
immediately before the quasiparticle peak vanishes \cite{Karsaki2008,Ganahl2015a,Lee2017}.
On the Mott insulating side, Refs.\ \onlinecite{Granath2014} and \onlinecite{Nishimoto2004}
show an extra peak or a shoulder feature on the inner side of the Hubbard bands,
similar to our findings.
In Fig.~\ref{fig:mott_u}
we also compare the $A(\omega)$ that is directly continued from the Green's function on the imaginary axis,
which shows a shoulder rather than a double peak.
While we hence cannot resolve within the maximum entropy uncertainty,
whether we actually have a shoulder or double peak structure,
it is clear that there is a feature in the upper Hubbard band.
Mathematically, this is necessitated by the strong scattering rate in this region.
A simple  physical picture or understanding of these side structures in the Hubbard bands is still missing.
Note, that also in strong coupling perturbation theory to second order such a shoulder
and hence asymmetry of the self-energy within the upper Hubbard band is observed  \cite{Kalinowski2002},
whereas the Hubbard-III approximation~\cite{Hubbard64} and the Falicov-Kimball 
model~\cite{vanDongen1997,Freericks2003} do not show such a shoulder. 
In agreement, Fig.~\ref{fig:mott_u} shows this feature for different values of $U$.
Since the scattering in BSE without $\mathbf k$ and with non-interacting
$A_0(\omega)$ is merely rescaled by $U^2$, it is clear from Fig.~\ref{fig:mott_u}
that the agreement of the position of the maximal scattering rate
between BSE and DMFT in  Fig.~\ref{fig:mott_u12_b5} (bottom, black-dotted vs.~red-dashed line) was by chance.

We can conclude that the one-electron scattering rate in a Mott insulator
is very different from an impact ionization picture. It is associated with the formation
($\omega\sim 0$) of the Hubbard bands and even side structures therein
($\omega\sim 4$ in  Fig.~\ref{fig:mott_u12_b5}).
The Hubbard bands are created by the interaction of the same electrons
we also use as a test charge for calculating the scattering rate.
If there is a local extra hole or electron, locally the  Hubbard bands deform.
Most noticeable this is in the filling of the Mott gap,
which does not only occur with increasing temperature \cite{Mo2004}
but also if we drive the system out of equilibrium \cite{Werner2014,Sorantin2018,wais2018}. 
If we have an extra electron in a disordered spin background of the DMFT Mott insulator,
it can hop or cannot hop to a neighboring site depending on the spin orientation of this neighbor. 
This leads to a large scattering rate without changing the number of double occupations.
These processes are included in the DMFT but not in the BSE,
they do not contribute to impact ionization (do not change the number of double occupations)
or major energy redistributions. 

\section{Conclusion}\label{sec:conclusion}
We have studied and compared scattering rates using two widely employed methods:
BSE and  DMFT. We have employed these methods out of their comfort zone,
where they cannot be applied with mathematical rigor. For DMFT this is the dimensionality of the systems studied (2D),
which is far away from the limit of infinite dimensions where DMFT become exact.
For the BSE it is the strong interaction regime of the Mott insulator,
where a rate equation with perturbatively determined scattering rates cannot safely be applied.
We have mitigated the latter in part by using the interacting spectral function
instead of the non-interacting DOS as the quasiparticle states whose occupation dynamics (here scattering rate) is calculated by BSE.

DMFT somewhat underestimates the scattering rates and by construction cannot
resolve their momentum-, only their energy-dependence.
This momentum dependence is particularly strong in the middle of the band
where the Van-Hove singularity is located. The physical reason behind both 
discrepancies is that the phase space for the scattering of a quasiparticle
with another quasiparticle explicitly depends on available unoccupied states
linked by momentum conservation. If we replace the momentum-conserving
$\delta$-function by a Gaussian with increasing width or directly disregard momentum conservation
in the BSE without $\mathbf k$, the scattering rates are reduced and the DMFT results
reproduced by BSE without $\mathbf k$  for the weakly correlated metal ($U=1$ or 2).

The biggest challenge for the BSE is the strongly interacting Mott insulating state.
Here the DOS is split into two Hubbard bands which we take as the starting quasiparticle
DOS in the BSE without $\mathbf k$. In the BSE, the scattering rate is due to impact ionization.
These processes are well described and in good agreement with DMFT~\cite{wais2018}.
However, in DMFT additional scattering processes which can be associated with the formation of the Hubbard bands
and shoulders therein dominate. The same specimen of electrons that through their interaction form the Hubbard bands
are also added as a charge probe, locally disturbing the spectrum.
These huge DMFT scattering rates are beyond a BSE description with a static DOS.

Scattering in an interacting  band insulator bears no similarity at all 
with that in the Mott insulator. It is strongly suppressed at low temperatures
since scattering is only possible if there are thermal excitations across the gap.
BSE without $\mathbf k$ and DMFT agree, while the BSE with momentum conservation has,
similar as for the weakly correlated metal,
somewhat larger scattering rates. The difference to the Mott insulator does
not only lie in the huge scattering associated with the Hubbard bands,
but also in the absence of impact ionization which dominates the scattering in BSE for a Mott insulator.
Impact ionization and Auger processes are only possible in a band insulator
through higher order in $U$ processes, through
quite small Coulomb matrix elements beyond the Kanamori interaction,
or a sizable hybridization between valence and conduction band.
This strongly suggests that Mott insulators are better suited than band insulators
for increasing the efficiency of solar cells through impact 
ionization \cite{Manousakis2010,Assmann2013,Werner2014,Sorantin2018,wais2018,Maislinger2020,Kauch2020a}.

\begin{acknowledgments}
We thank  M. Eckstein and P. Werner for discussions, and acknowledged financial support  from the Austrian Science Fund (FWF)  through  the Doctoral  School  W1243  Solids4Fun  (Building Solids for Function; MW) and  project P30997 (M.W., J.K., K.H.), and from   Nanyang Technological University, NAP-SUG (M.B.). Calculations have been done in part on the Vienna Scientific Cluster (VSC).
\end{acknowledgments}

\appendix

\section{Scattering rate from the retarded Green's function}\label{app:gft}
\subsection{Derivation of the formula}
If we linearize the self-energy around the (real part of the) pole 
$\tilde{\epsilon}(\mathbf{k})=Z[{\epsilon}(\mathbf{k})+\mathrm{Re}\Sigma_R(0)-\mu]$
we get Eq.~(\ref{eq:G_QP})
whose Fourier transformation is
\begin{equation}
  \label{eq:gft-w-t}
  G(\mathbf{k}, t) = \frac{1}{2\pi} \int_{-\infty}^\infty d\omega 
  \frac{Z e^{-i\omega t}}{\omega -  \tilde{\epsilon}(\mathbf{k}) - Z \mathrm{Im}\Sigma_R(\omega)}.
\end{equation}
Note that a linearization of the self-energy around $\omega=0$ might not be justified any longer 
if the pole is at large frequencies and that there may be more than one pole at a given $\mathbf k$. 
For example for the Mott insulator we have two poles for each $\mathbf k$. 
However this merely means that we have $\mathrm{Re}\Sigma_R(\tilde{\epsilon})$ 
instead of $\mathrm{Re}\Sigma_R(0)$ and that we have a sum of poles in Eq.~(\ref{eq:gft-w-t}) 
instead of a single one, respectively. 
With these modifications, we can use the same procedure as discussed in the following for a single pole.

The integral (\ref{eq:gft-w-t}) can be solved by closing the contour on the lower complex half-plane of frequencies 
(since $t\!>\!0$ the integrand is exponentially suppressed here). 
Then, Eq.\ \eqref{eq:gft-w-t} can be computed by the residue theorem.
The  pole is at 
$[{\rm Re} \omega_p,{\rm \Im}\omega_p]=[\tilde{\epsilon}(\mathbf{k}),Z{\rm \Im}\Sigma(\omega_p\approx \tilde{\epsilon}(\mathbf{k}))]$.
Here, in principle ${\rm \Im}\omega_p$ would have to be obtained self-consistently, 
but if ${\rm \Im}\Sigma$  is small we can (approximately) calculate it using only the real part the pole.
The residue theorem then yields for the integral in Eq.~(\ref{eq:gft-w-t})
\begin{equation}
  \label{eq:residue-contribution}
  2 \pi i \lim_{\omega\rightarrow{\omega}_p} (\omega-{\omega}_p) 
  \frac{Z e^{-i\omega t}}{\omega - \tilde{\epsilon}(\mathbf{k}) -Z{\rm Im}\Sigma_R(\tilde{\epsilon})}
  \propto e^{-it\tilde{\epsilon}} e^{t Z\mathrm{Im}\Sigma_R(\tilde{\epsilon})};
\end{equation}
or for the probability to find 
a particle that is added at time $0$ to the quasiparticle state $\mathbf k$ still in this state at a later time $t$:
\begin{equation}
  \label{eq:approximate-decay}
  |G_R(\mathbf{k}, t)|^2 \propto e^{2 Z \mathrm{Im} \Sigma_R(\tilde{\epsilon}(\mathbf{k}))t},
\end{equation}
which yields the (inverse) life time Eq.~(\ref{eq:scatrat-dmft-w}).

\subsection{Analytic example}
\begin{figure}
  \includegraphics[width=0.4\textwidth]{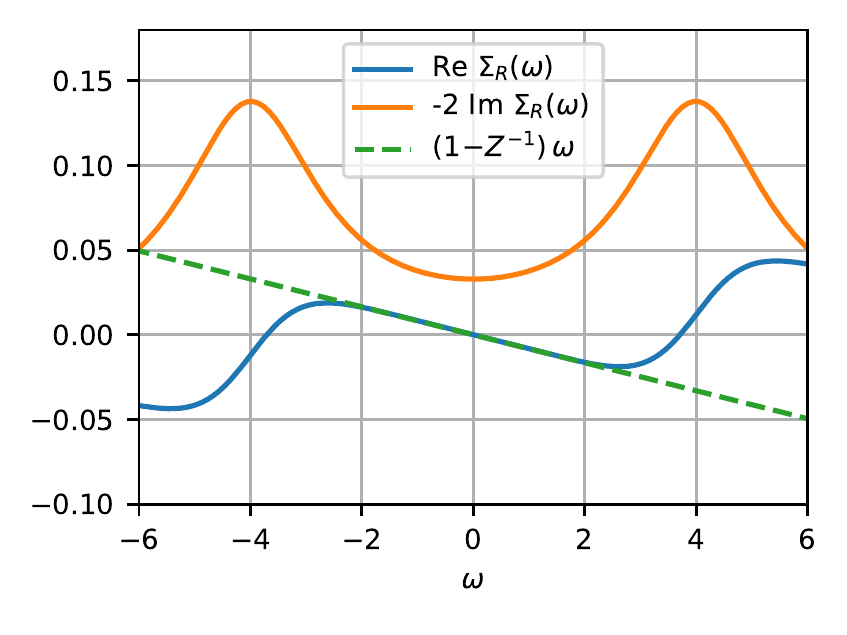}
  \caption{\label{fig:model-se} Model self-energy of Eq.\ \eqref{eq:model-self-energy} for $A=0.1$.}
\end{figure}
It is instructive to consider an example, where everything can be
computed exactly, such that we can test the above approximations.
We consider a self-energy of the form
\begin{equation}
  \label{eq:model-self-energy}
  \Sigma(\omega) = A \Big( \frac{1}{\omega-E+i\alpha} + \frac{1}{\omega+E+i\alpha} \Big).
\end{equation}
For parameters $A=0.1$, $E=4$, $\alpha=1.5$ it is very similar in size and shape
to our results for the single-orbital weak coupling results for $\beta\approx 2$,
as shown in Fig.\ \ref{fig:model-se}.

\begin{figure}
  \includegraphics[width=0.4\textwidth]{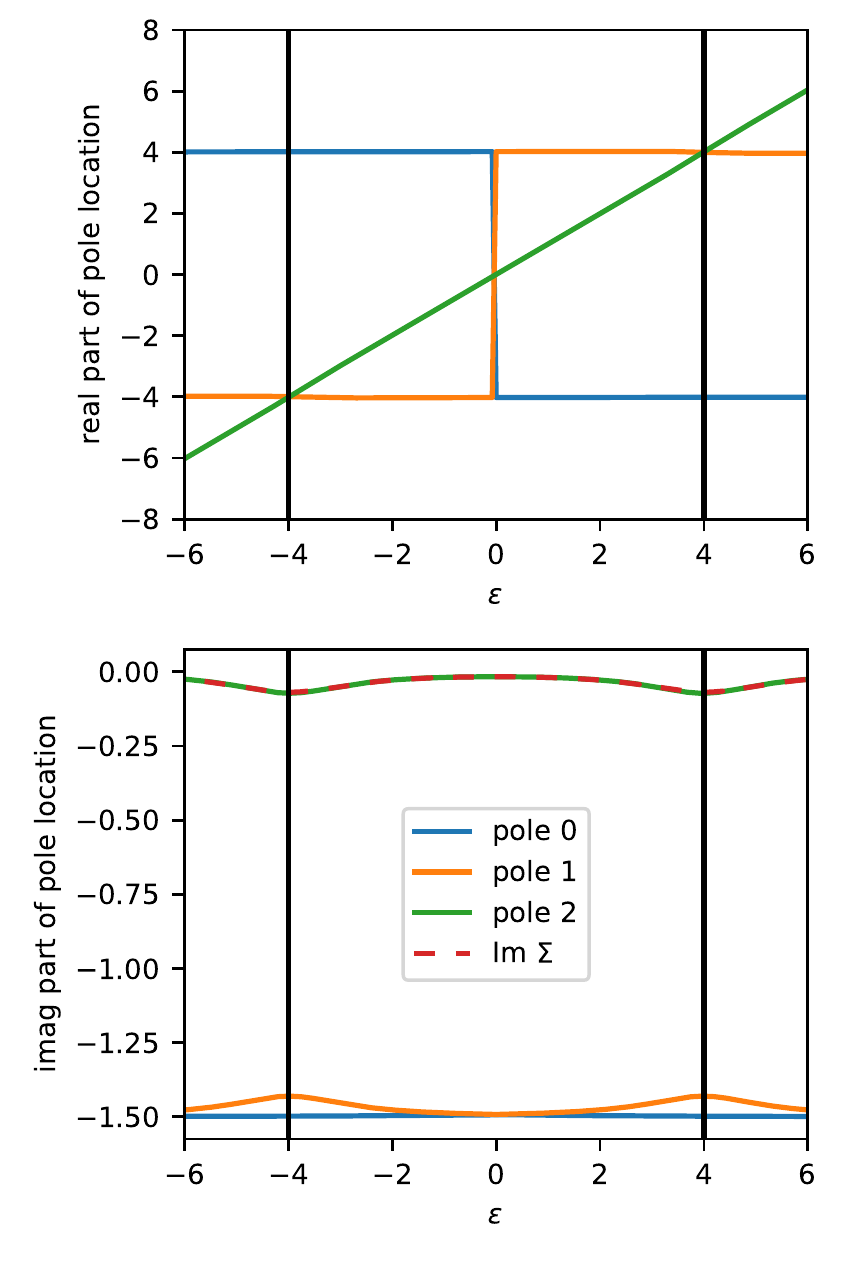}
  \vspace{-1em}
  \caption{\label{fig:pole-locations-small}Pole locations with model self-energy. Clearly the imaginary part of the
  location of the pole with the smallest imaginary part (pole 2) agrees well with the imaginary
  part of the self-energy ${\rm \Im} \Sigma (\omega=\tilde\epsilon)$. The imaginary parts of the other two poles are much larger.
  $Z\approx 0.99$ for this case.}
\end{figure}

\begin{figure}
  \includegraphics[width=0.4\textwidth]{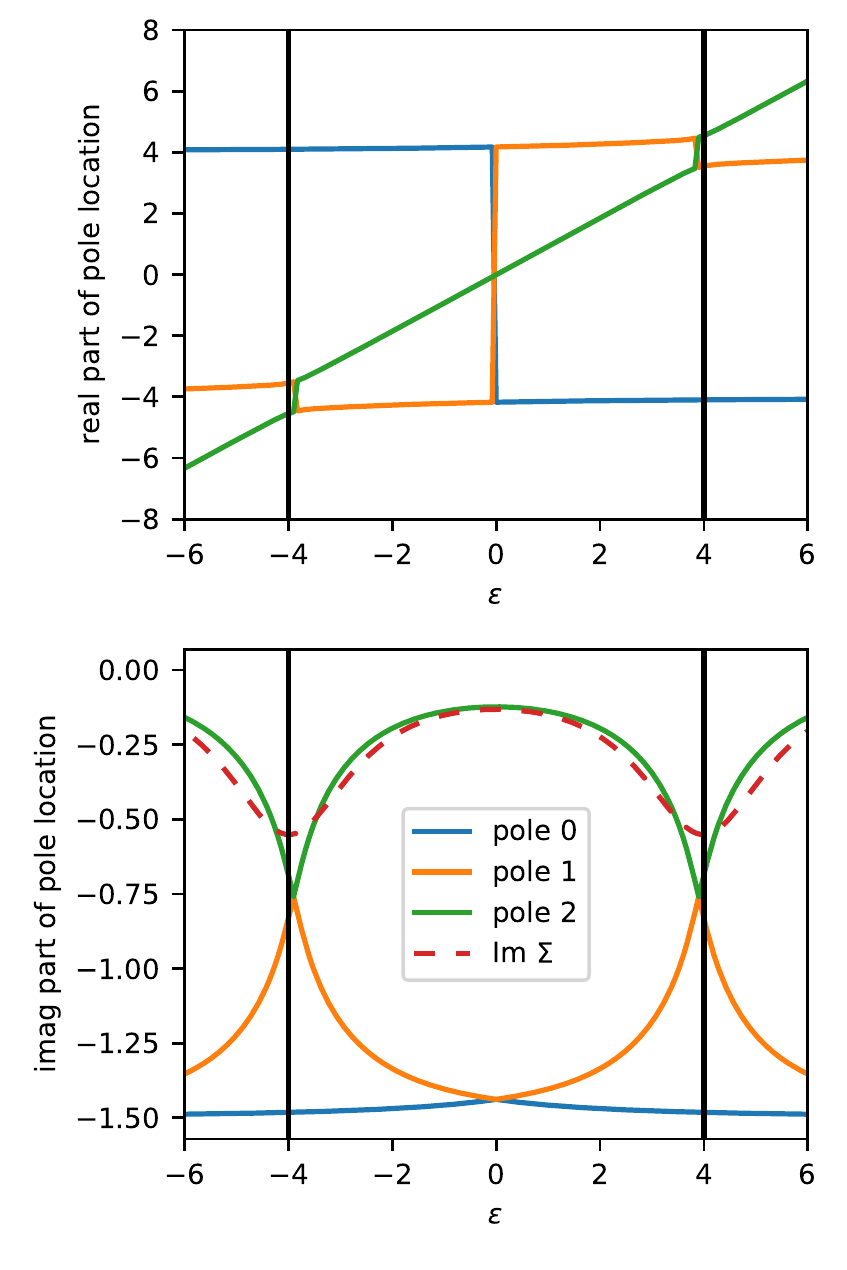}
  \vspace{-1em}
  \caption{\label{fig:pole-locations-large} Same as Fig.~\ref{fig:pole-locations-small} but with $A$ increased by a factor of 8.
  Also here the imaginary part of the location of the pole with the smallest imaginary part
  is similar to the imaginary part of the self-energy, but not in quantitative agreement.
  At the band edge, a second pole is of similar size and will thus have considerable influence
  on the scattering rate. $Z\approx 0.94$ for this case.}
\end{figure}

If we insert this self-energy together with $\mu=0$ 
into the Green's function Eq.~(\ref{eq:g-ret-w-explicit}), the locations of the
poles are determined by a cubic equation in $\omega$.
It is possible to solve this equation analytically for arbitrary parameters.
Notably, the pole locations will depend on $\epsilon(\mathbf{k})$.
The locations of the poles in dependence of $\epsilon$ are shown in Fig.\ \ref{fig:pole-locations-small},
and for a larger value of $A=0.8$ in Fig.\ \ref{fig:pole-locations-large}. 
Pole 2 is the one, we usually associate with the quasiparticle.

\begin{figure*}[tb]
  \begin{minipage}{14cm}
  \includegraphics[width=14cm]{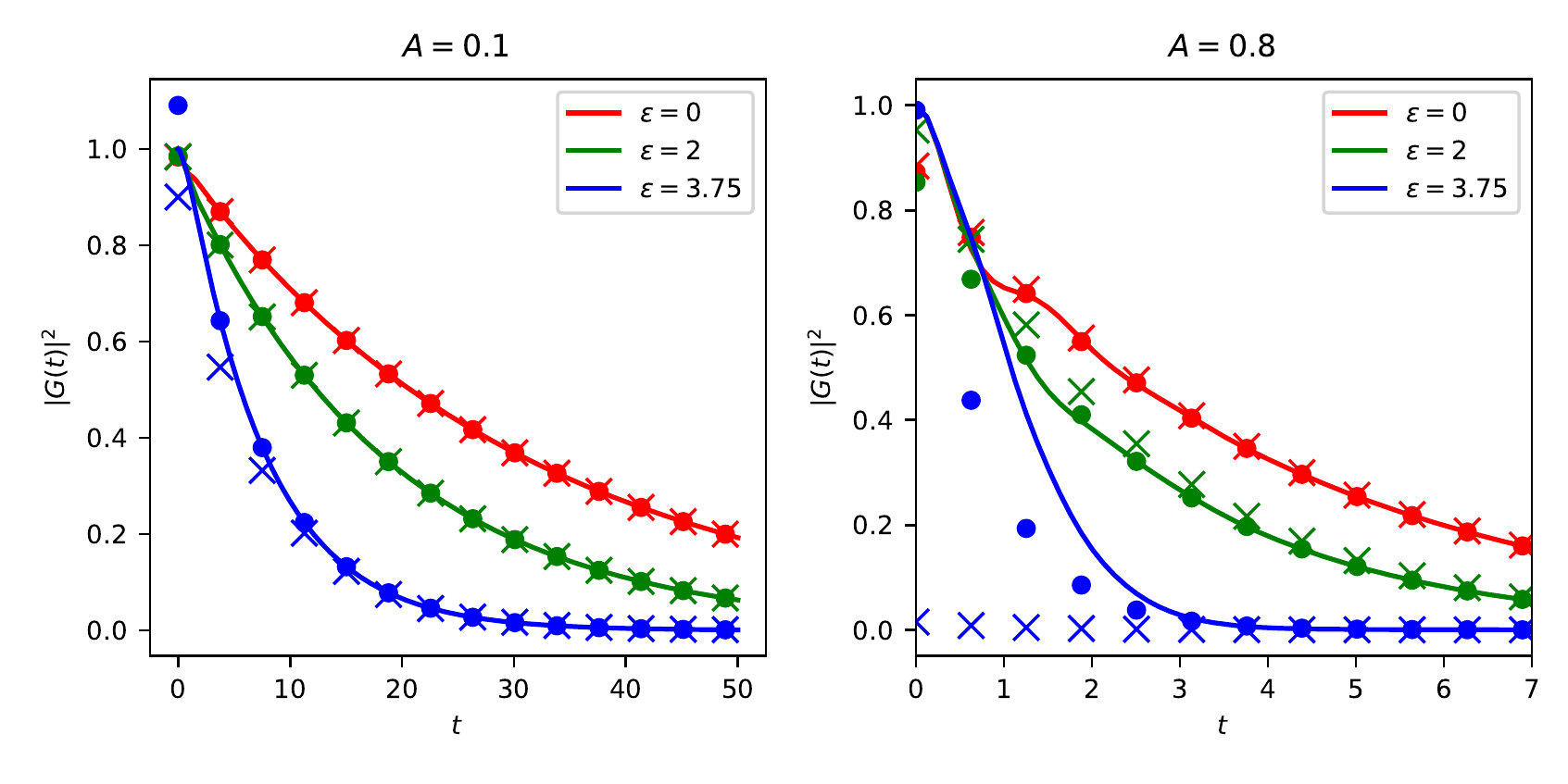}
  \end{minipage}\hfill
  \begin{minipage}{3.8cm}
  \caption{\label{fig:gt}Absolute square of the time-dependent Green's function for the model self-energy with  $A=0.1$ (left) and $A=0.8$ (right).  Solid lines are time-dependent Green's
  functions that are computed by Eq.\ \eqref{eq:gft-w-t}. Dots show the contribution of the residue
  at the pole with smallest $|\mathrm{Im}\hat{\omega}|$. Crosses show the approximation given by
  Eq.\ \eqref{eq:approximate-decay}.}
  \end{minipage}
\end{figure*}

Now it is possible to compute the time-dependent Green's function exactly by evaluating the
Fourier integral in Eq.\ \eqref{eq:gft-w-t}. The results for a few different values of ${\epsilon}$
are shown as solid lines in Fig.\ \ref{fig:gt}. 
Given the exact $G(t)$ as a reference, we show the contribution of the residue of the pole (pole 2)
that is closest to the real axis, Eq.\ \eqref{eq:residue-contribution} as dashed lines.
Finally we also show the exponential decay where the scattering rate was approximated by
the imaginary part of the self-energy. Clearly, for small values of $\epsilon$ this
is an excellent approximation. Closer to the band edge it does not match so well any more,
especially in the case of the large self-energy (large $A$; right panel). 
This mismatch may already be anticipated when looking at Fig.\ \ref{fig:pole-locations-large}.

\section{Convolution Method}\label{chap:appConvolution}
For the case of a single band at half-filling, we here reformulate the expression for the 
scattering rate in BSE without $\bold k$, Eq.~\eqref{eq:boltznok}, to gain some further analytical insight.
To this end, we define the particle density as $n_p (\omega) \equiv f_\textrm{FD} (\omega) A_0(\omega)$
and the hole-density as $n_h (\omega) \equiv (1-f_\textrm{FD} (\omega)) A_0(\omega)$. 
With these definitions and the scattering amplitude for the one-band system introduced in Section~\ref{chap:weak1band}
we may rewrite Eq.~\eqref{eq:boltznok} as
\begin{equation}
\begin{split}
&\frac{1}{\tau_n(\epsilon_0)} = 2 \pi U^2\int \mathrm d \epsilon_1 \mathrm d \epsilon_2 \mathrm d \epsilon_3 \Big [ \delta(\epsilon_0 + \epsilon_1 - \epsilon_2 - \epsilon_3)\\
&\quad   \times \Big ( n_h(\epsilon_1)n_p(\epsilon_2) n_p(\epsilon_3) + n_p(\epsilon_{1}) n_h(\epsilon_2) n_h(\epsilon_3) \Big )\Big ] .\\
\end{split} \label{eq:boltznok2}\!
\end{equation}
For a system with particle-hole symmetry it holds that $n_p(\omega) = n_h(-\omega)$. 
Using this property and the definition of the convolution 
$(a * b)(\omega) \equiv \int \mathrm d \tilde  \omega~ a(\omega - \tilde \omega) b(\tilde \omega)$,
Eq.~\eqref{eq:boltznok2} can be further reduced to 
\begin{equation}
\frac{1}{\tau_n(\omega)} = 2 \pi U^2 \big [ g(\omega) +  g(-\omega) \big ] , \label{eq:twopeak}
\end{equation}
with 
\begin{equation}
g(\omega) \equiv (n_p * n_p * n_p)(\omega)  \textrm{ .}
\end{equation}
The above equation states that the scattering rate consists of the sum of the particle density
convoluted with itself twice, and its mirrored version.

\begin{figure}
 \includegraphics[width=6.5cm]{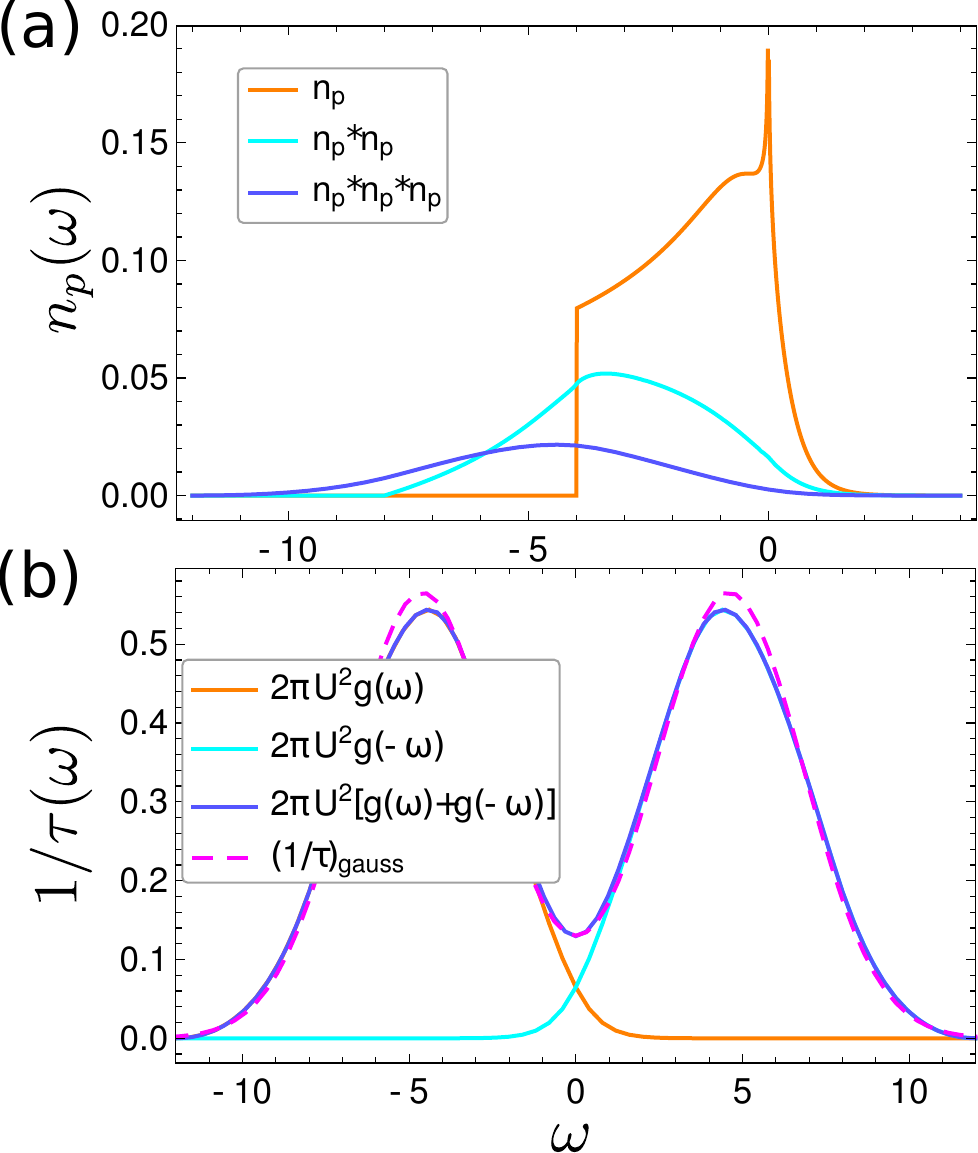}
 \caption{(a) Particle-density $n_p$ and the particle-density convoluted one- and two-times with itself. (b) Scattering rate in BSE without $\mathbf k$  calculated with the convolution method compared to the approximated scattering rate that consists of two Gaussian functions.}\label{fig:convMeth}
\end{figure}

According to the central limit theorem \cite{Montgomery2018}, 
a function with compact support becomes a Gaussian function 
in the limit when it is convoluted an infinite times with itself. 
If the particle-density is smooth, 
the result after two convolutions with itself is already very similar
to a Gaussian, see Fig.~\ref{fig:convMeth} (a). 

This allows us to further reduce complexity and increase understanding: 
A general Gaussian function, i.e.
\begin{equation}
f_\textrm{gauss}(\omega) \equiv \frac{\alpha}{\sigma \sqrt{2 \pi}} \times \exp[-\frac{{(\omega - \omega_0)}^2}{2 \sigma ^2}] 
\end{equation}
is completely defined by three parameters: 
its integral value $\alpha$, its variance $\sigma^2$ and its zero-point $\omega_0$. 
For a given Gaussian these three parameters can be calculated from its zeroth-, first- and second-moment,
\begin{align}
\alpha &= F_0[f_\textrm{gauss}] , \label{eq:alpha} \\ 
\omega_0 &= \frac{F_1[f_\textrm{gauss}]}{F_0[f_\textrm{gauss}]} ,\\
\sigma ^2&= { \frac{F_2[f_\textrm{gauss}]}{F_0[f_\textrm{gauss}]} - {\left ( \frac{F_1[f_\textrm{gauss}]}{F_0[f_\textrm{gauss}]} \right )}^2 }, \label{eq:sigma}
\end{align}
where $F_n[f]$ is the $n$-th moment of the function $f(\omega)$, i.e.,
\begin{equation}
F_n[f ] \equiv \int \mathrm d \omega f(\omega) {\omega}^n \textrm{ .}
\end{equation}

We can now approximate the function $g(\omega)$ with a Gaussian function 
by calculating the parameters $\alpha$, $\omega_0$ and $\sigma$ from its moments 
using Eqs.~\eqref{eq:alpha}-\eqref{eq:sigma} with $g$ instead of $f_\textrm{gauss}$. 
The function $g(\omega)$ is calculated from a convolution of the particle-density, 
hence, the moments of $g(\omega)$ can be directly inferred from the moments of the particle-density $n_p(\omega)$,
\begin{align}
F_0[g] = & { \left ( F_0[n_p] \right ) }^3 , \label{eq:mom0}\\
F_1[g] =& 3 F_1[n_p] { \left ( F_0[n_p] \right ) }^2 , \\
F_2[g] =& 3 \left [ F_2[n_p] { \left ( F_0[n_p] \right ) }^2 + 2 {\left ( F_1[n_p] \right ) }^2 F_0[n_p] \right ] . \label{eq:mom2}
\end{align}
With the Eqs.~\eqref{eq:alpha}-\eqref{eq:sigma} and the Eqs.~\eqref{eq:mom0}-\eqref{eq:mom2} 
the parameters for the Gaussian function can eventually be calculated from the moments of particle-density as 
\begin{align}
\alpha &= { \left ( F_0[n_p] \right ) }^3 , \label{eq:alpha2} \\ 
\omega_0 &= 3 \frac{F_1[n_p]}{F_0[n_p]} ,\\
\sigma^2 &= 3 \left [{ \frac{F_2[n_p]}{F_0[n_p]} - {\left ( \frac{F_1[n_p]}{F_0[n_p]} \right )}^2 } \right ], \label{eq:sigma2}
\end{align}
and the scattering rate may be approximated as
\begin{equation}
  \label{eq:scatrat-conv}
\left ( \frac{1}{\tau(\omega)} \right )_\textrm{gauss} \equiv 2 \pi U^2 \left [ f_\textrm{gauss}(\omega) + f_\textrm{gauss}(-\omega) \right ] .
\end{equation}
 Fig.~\ref{fig:convMeth} (b) shows that this approximate Gaussian agrees with the exact BSE without $\mathbf k$ scattering rate to very good approximation. This explains the origin of the maximal scattering rate and why this maximum lies within the bandwidth of the DOS not at its edge as in the BSE with $\mathbf k$.

 \section{Connection to 2$^\text{nd}$ order perturbation theory}
 \label{sec:IPT}
The scattering rate of Eq.\ \eqref{eq:boltznok2} is actually equivalent to second order iterated
perturbation theory (IPT) \cite{Georges1992a,Kajueter96,Kajueter1996b}. In IPT, the DMFT self-energy is calculated in second order in $U$ from  the impurity Green's function $\cal G$. Directly on the real frequency axis and in terms of the impurity spectral function ${\cal A}(\omega)=-1/\pi \; {\rm Im} {\cal G}(\omega)$ the IPT self-energy reads  (see e.g.\ Eq.~(22) in \cite{Potthoff1997b} where $\epsilon_1$ and $\epsilon_2$ are exchanged)
\begin{eqnarray}
 \Sigma_R(\omega)&&  = \lim_{\alpha\rightarrow0} U^2\int \mathrm d \epsilon_1 \mathrm d \epsilon_2 \mathrm d \epsilon_3 {\cal A}(\epsilon_1)  {\cal A}(\epsilon_2) {\cal A}(\epsilon_3) \\
  &&\!\!\! \times \frac{f_{\rm FD}(\!-\!\epsilon_1\!)f_{\rm FD}(\epsilon_2\!)f_{\rm FD}(\epsilon_3\!)\!+\!f_{\rm FD}(\epsilon_1\!)f_{\rm \rm FD}(\!-\!\epsilon_2\!)f_{\rm FD}(\!-\!\epsilon_3\!)}{\omega+i\alpha+\epsilon_1-\epsilon_2-\epsilon_3} . \nonumber
\end{eqnarray}
With $f_{\rm FD}(-\epsilon)=1-f_{\rm FD}(\epsilon)$ and $\lim_{\alpha\rightarrow0} {\rm Im} 1/(\omega+\epsilon_1-\epsilon_2-\epsilon_3)=-\pi \delta (\omega+\epsilon_1-\epsilon_2-\epsilon_3)$ this yields Eq.~\eqref{eq:boltznok2} or Eq.~\eqref{eq:boltznok} if we replace  ${\cal A}(\epsilon)$ by  ${A}_0(\epsilon)$  which is possible to lowest order in $U$ or the first iteration of the IPT. Through the DMFT self-consistency condition~\cite{Georges1992,Georges1996}  ${\cal G}(\omega)^{-1} =G(\omega)^{-1}+\Sigma(\omega)$, ${\cal A}(\epsilon)$ is updated in subsequent iterations.

\section{Simplifications for the two-orbital case due to particle-hole symmetry}
\label{chap:simp2band}
In this Section, we discuss some simplifications that are possible  due to
particle-hole symmetry and a density-density interaction $U$ which is the same for all orbitals. 
Indeed, the scattering rate can be actually calculated from a single band in BSE. The reason for this is as follows:
For the BSE with momentum conservation the scattering rate in the upper band reads
\begin{equation}
\begin{split}
&\frac{1 }{\tau_{B}(\bold k_0)} = 6 \pi U^2\frac{1}{{V_{BZ}}^2} \int \mathrm d^2 k_1 \mathrm d^2 k_2 \mathrm d^2  k_3 \Big [  \\
&\times \delta( \epsilon_{B}(\bold k_0) + \epsilon_{B}(\bold k_1) - \epsilon_{B}(\bold k_2) - \epsilon_{B}(\bold k_3))\\
&\times \sum_{\bold G} \delta ( \bold k_0 + \bold k_1 - \bold k_2- \bold k_3 + \bold G)\\
&  \times   \Big ( (1-f_\textrm{FD}(\epsilon_{B}(\bold k_1)) )f_\textrm{FD}(\epsilon_{B}(\bold k_2)) f_\textrm{FD}(\epsilon_{B}(\bold k_3))\\
&  + f_\textrm{FD}(\epsilon_{B}(\bold k_1)) (1- f_\textrm{FD}(\epsilon_{B}(\bold k_2)))(1- f_\textrm{FD}(\epsilon_{B}(\bold k_3))) \Big )\Big ],
\end{split} \label{eq:boltzwithkRed}
\end{equation}
and for the BSE without $\bold k$ case it reads
\begin{equation}
\begin{split}
&\frac{1}{\tau_B(\epsilon_0)} = 6 \pi U^2 \int \mathrm d \epsilon_1 \mathrm d \epsilon_2 \mathrm d \epsilon_3 \Big [ \\
& \quad \times \delta(\epsilon_0 + \epsilon_1 - \epsilon_2 - \epsilon_3) A_0^B(\epsilon_1) A_0^B(\epsilon_2) A_0^B(\epsilon_3)\\
&\quad \quad \quad \times \Big ( (1-f_\textrm{FD}(\epsilon_{1}) )f_\textrm{FD}(\epsilon_{2}) f_\textrm{FD}(\epsilon_{3})\\
& \quad\quad \quad + f_\textrm{FD}(\epsilon_{1}) (1- f_\textrm{FD}(\epsilon_{2}))(1- f_\textrm{FD}(\epsilon_{3})) \Big )\Big ] .
\end{split} \label{eq:boltznokRed}
\end{equation}
Due to particle-hole symmetry it further holds that $1/\tau^A(-\omega) = 1/\tau^B (\omega) \equiv 1/\tau (\omega)$. 
The multiplicative factor of 3 compared to the one-band case (Section~\ref{chap:weak1band}) in the scattering amplitude reflects the different scattering processes an electron in the upper band may perform: an electron with a certain spin $\sigma$ in band $B$ may scatter with an an electron $B \bar \sigma$, $A \sigma$ and $A \bar \sigma$. Since the density-density interaction does not allow for spin-flips and  pair-hopping nor impact excitation which would require
an interaction of the form $c^{\dagger}_{i B  \bar \sigma} c^{\phantom{dagger}}_{i A \bar \sigma}
c^{\dagger}_{i B \sigma } c^{\phantom{dagger}}_{i B \sigma }$ nor Auger excitations,
there are no further allowed processes to be taken into account.
Since the interaction between the bands is the same as within the bands ($U=V$), 
all scattering processes have the same scattering amplitude $\propto 2 \pi U^2$ ($\propto 2 \pi \frac{1}{ {V_{BZ}}^2 }U^2$),
eventually leading to $\propto 3 \times 2 \pi U^2 = 6 \pi U^2$ ($\propto 6 \pi \frac{1}{ {V_{BZ}}^2 }U^2$).

\section{Band insulator with renormalized bands} \label{chap:twoBandInteracting}
\begin{figure*}[tb]
  \begin{minipage}{12cm}
    \includegraphics[width=12cm]{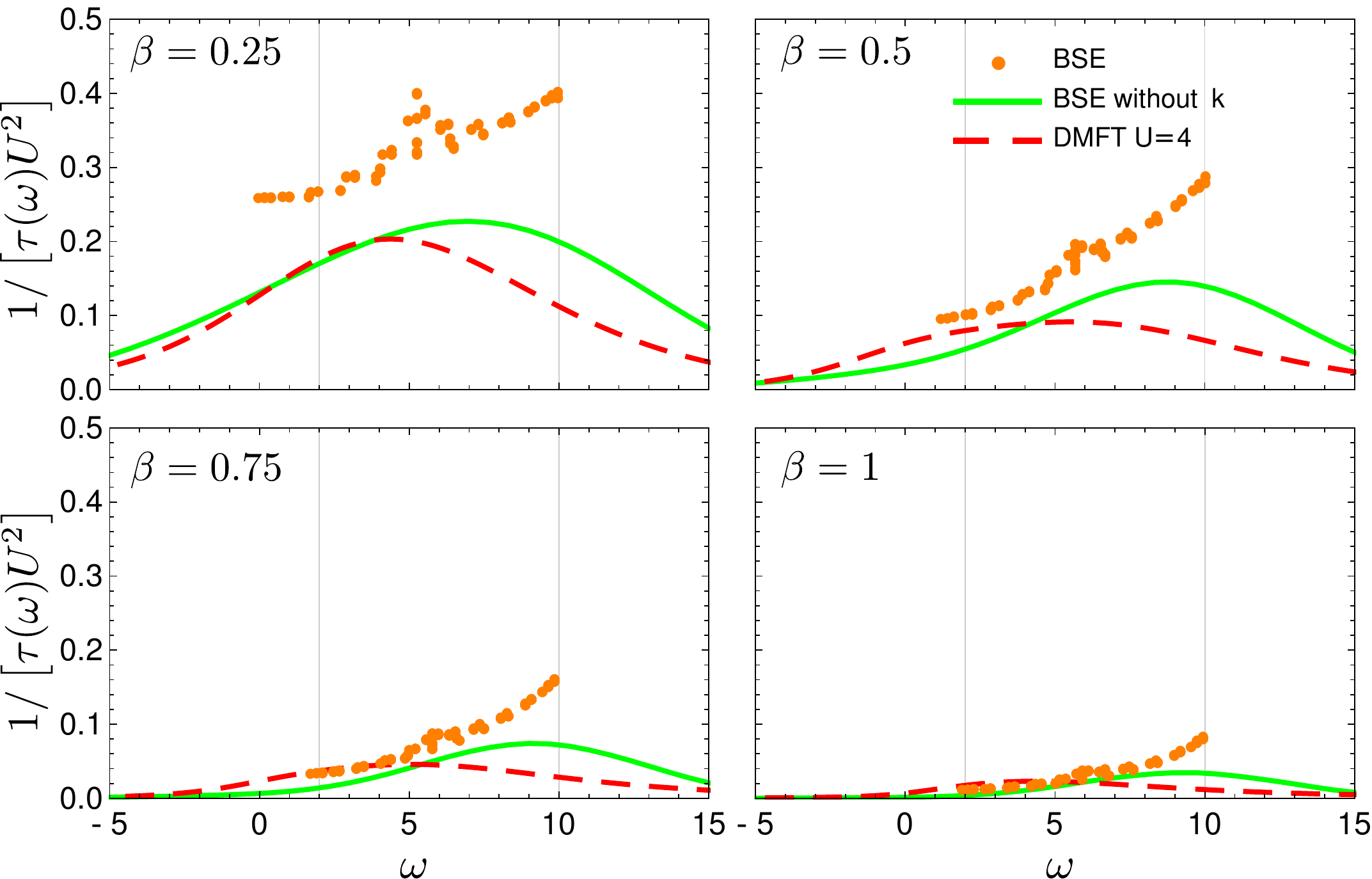}
    \end{minipage}\hfill
 \begin{minipage}{5cm}
 \caption{Scattering rate normalized by $U^2$ for the two-band case with $U=4$. 
   Same as Fig.~\ref{fig:2band} but using the interacting DMFT spectral density  $A(\omega)$ instead of  $A_0(\omega)$ in Eqs.~(\ref{eq:A0A}),~(\ref{eq:A0B}) for the BSE without $\mathbf k$. For the BSE with momentum conservation the dispersion and gap has been adapted to the DMFT as well. Using the interacting spectral  function leads to a better agreement with DMFT.}\label{fig:2bandinteracting}\vfill
 \end{minipage}
\end{figure*}
As already mentioned in Section~\ref{weakCoupling2Band} the deviation of BSE without $\bold k$ and DMFT 
can be reduced by using the interacting spectral density instead of the non-interacting DOS
[$A_0(\omega) \to A(\omega)$ in Eq.~\eqref{eq:boltznokRed}]. 
For the theoretical justification and background of this procedure, see  Ref.~\cite{wais2018}.

For the band insulator, the main difference between $A_0(\omega)$ and $A(\omega)$ is that thermal excitation across the gap reduce the difference in Hartree energy and hence the band gap. Moreover,  with more thermal excitations there are more particles an electron or hole can scatter with. Hence the scattering rate is
enhanced, which in turn leads to broadening effects in  $A(\omega)$, visible in Fig.~\ref{fig:2bandSpecDens} at the band edges and the van Hove singularity.

In order to make all methods comparable, we are also using a renormalized band
for the BSE with momentum conservation.
For that purpose we calculate the momentum-dependent spectral density $A_{\bold k}(\omega)$
and then extract for each given momentum the corresponding energy where the spectral density has its maximum.
In this way one can obtain a renormalized dispersion relation which we then use for the calculation of the scattering rate.

Fig.~\ref{fig:2bandinteracting} shows the thus obtained BSE results at $U=4$
with and without momentum conservation; the DMFT result is the same as in
Fig.~\ref{fig:2band}. Using the interacting spectral function and correspondingly renormalized bandstructure,
reduces the scattering rate in the middle of the upper band at $\omega\sim 6$.
In contrast the scattering rate at small $\omega\sim 0$ is enhanced at high temperatures.
Here, the interacting spectrum has states in the (pseudo)gap.
Overall, using the interacting DOS as a starting point the BSE can explain most changes
of the $U=4$ DMFT scattering rate compared to $U=2$.


%

\end{document}